\documentclass{aa}  

\usepackage{graphicx}
\usepackage{txfonts}
\usepackage{lipsum}
\usepackage{pifont}
\usepackage[colorlinks=true,linkcolor=blue, filecolor=blue, urlcolor=blue, citecolor=blue]{hyperref}

\usepackage{subcaption}         
\usepackage{lscape}             
\usepackage{placeins}           
\usepackage{tikz, hyperref}

\usepackage{orcidlink}

\defcitealias{MatthijsseInprepA&APaperII}{Paper II}
\defcitealias{Lin2021ApJ...907...64L}{L21}
\defcitealias{Sengupta2023ApJ...942...74S}{SU23}

\begin{document}

   \title{Dust-driven vertical shear instability in an isothermal shearing box}
   \subtitle{I. Turbulence and gas dynamics without dust feedback}

   \author{Jip Matthijsse\orcidlink{0009-0001-5298-9725}\thanks{Corresponding author; j.p.matthijsse@tudelft.nl}
          \inst{1}
          \and
          Hossam Aly\orcidlink{0000-0002-1342-1694}\inst{1}
          \and
          Sijme-Jan Paardekooper\orcidlink{0000-0002-8378-7608}\inst{1}
          \and
           Richard P. Nelson\orcidlink{0000-0002-9687-8779}\inst{2}}

   \institute{Planetary Exploration, Technical University Delft, Kluyverweg 1 2629 HS Delft, The Netherlands
         \and
         Astronomy Unit, Dept. of Physics and Astronomy, Queen Mary University of London, London E1 4NS, UK
             }

 \date{Received XXXX; accepted YYYY}

 
  \abstract
    {The “classical” vertical shear instability (VSI) is a baroclinic instability that arises from height-dependent azimuthal velocities and is a leading candidate for driving anisotropic turbulence in discs with extended vertical modes. The “classical” VSI requires fast cooling times and radial temperature gradients, which are absent in an isothermal stratified (barotropic) setup.}
    {We investigate the dust-driven vertical shear instability in an isothermal, stratified shearing box, where the VSI is generated by drag forces acting on the gas due to dust rather than radial thermal stratification.}
    {We construct two-fluid (gas+dust) vertically stratified equilibria including dust diffusion and use these profiles to initialise 2D (radial–vertical) hydrodynamical simulations with FARGO3D. To separate dust-driven vertical shear instability (DVSI) from drag instabilities, we employ a “dust-analogue” model in which dust-induced drag on the gas is prescribed as a time-invariant and height-dependent acceleration, such that the underlying spatial distribution of the dust is assumed to be constant.}
    {The DVSI grows from regions of strongest vertical shear away from the midplane, exciting predominantly radially short, vertically extended disturbances (large $k_x/k_z$) and producing banded structures in the perturbed azimuthal and vertical velocities during the linear phase. In the nonlinear regime, the instability saturates via a Kelvin–Helmholtz-like parasitic instability that breaks the coherent vertical-shear modes into smaller-scale eddies, which balance the growth of the vertically extended modes, driving sustained vertical stirring. }
    {Dust-driven vertical shear can drive anisotropic turbulence and vertical mixing in an isothermal model without invoking global thermal gradients and could play an important role in the stratified streaming instability.}


   \maketitle


\section{Introduction}

\par In protoplanetary discs (PPDS), turbulence plays an important role in their evolution and planet formation. The relative velocities between dust particles are driven by gas turbulence, which affects dust evolution through collisions \citep[see e.g.][]{Birnstiel2012A&A...539A.148B}. Turbulence also influences the formation of planetesimals by stirring the dust and can both concentrate and disperse it \citep[][]{Johansen2007Natur.448.1022J}. In addition, turbulent diffusion counteracts the vertical settling of dust particles: smaller grains remain more vertically extended and can be transported upward through vertical mixing \citep[][]{Dubrulle1995Icar..114..237D, Youdin2007Icar..192..588Y}. This vertical redistribution of dust can significantly affect the disc chemistry \citep[see e.g.][]{Semenov2006ApJ...647L..57S,semenov2011ApJS..196...25S, Heinzeller2011ApJ...731..115H, Woitke2022A&A...668A.164W, Ligterink2024NatAs...8.1257L}.

\par A range of dynamical instabilities can drive turbulence in PPDs. If the disc is sufficiently ionised, the magnetorotational instability \citep[MRI;][]{Balbus1991ApJ...376..214B}  is considered a leading mechanism for driving turbulence. However, PPDs are often poorly ionised, and at larger radii there are regions where magnetically driven instabilities are inefficient to drive turbulence \citep[][]{Gammie1996ApJ...457..355G}.

\par In these so-called dead zones, turbulence is thought to be driven predominantly by hydrodynamic instabilities, such as the Vertical Shear Instability \citep[VSI;][]{Urpin1998MNRAS.294..399U, Urpin2003A&A...404..397U, Nelson2013MNRAS.435.2610N} \citep[for an overview see][]{Lesur2023ASPC..534..465L}. Here, the “classical” VSI is driven by height-dependent azimuthal velocity that arises from the radial temperature gradient in the disc. If there is no radial temperature structure, we would need an alternative that could drive the vertical shear profile.

\par A vertical dependence of the gas azimuthal velocity can arise through a settled dust layer \citep[][hereafter \citetalias{Lin2021ApJ...907...64L}]{Lin2021ApJ...907...64L}. The pressure-supported gas orbits sub-Keplerian, while the dust orbits Keplerian. This velocity difference leads to a drag force that affects the equilibrium gas and dust velocities \citep[][]{Nakagawa1986Icar...67..375N}. The equilibrium gas velocity depends on the dust-to-gas ratio and the particle size. In a stratified disc, the dust is concentrated around the mid-plane due to settling \citep[][]{Dubrulle1995Icar..114..237D}. As a result, the dust-to-gas ratio varies with height z, and the equilibrium gas velocity becomes vertically dependent.

\par In this work, we investigate the dust-driven vertical shear instability (DVSI) in a stratified, isothermal shearing box. We investigate the linear growth and saturated regime of the instability using numerical simulations.

\par The differences between the “classical” VSI and the DVSI are highlighted in Sect.~\ref{sec:VSI}. The governing equations and the equilibrium profiles for the gas and dust velocities and densities are derived in Sect.~\ref{sec:eq_of_motions}. The numerical setup and the list of simulations are described in Sect.~\ref{sec:numerical_aprouch}. In Sect.~\ref{sec:DVSI} we investigate the evolution of the dust-driven vertical shear instability (DVSI), including its linear growth phase, nonlinear development and convergence with grid resolution. The dependence of the instability on the main physical parameters is explored in Sect.~\ref{sec:parameterstudy}. Finally, the results are discussed and summarised in Sect.~\ref{sec:discussion_conclusion}.


\section{The vertical shear instability}\label{sec:VSI}

The free energy that drives the VSI comes from vertical shear in the PPD. The angular velocity depends on both radius and height ($R, z$), which can lead to the growth of axisymmetric adiabatic perturbations when the Solberg–Høiland stability criteria is violated \citep[see][]{Nelson2013MNRAS.435.2610N}. These criteria are
\begin{align}
        \kappa^2 + N_R^2+N_z^2 >& \, 0,\\
        -\partial_R P\left( \kappa^2 \partial_z S-R\partial_z \Omega^2\partial_R S  \right) >& \, 0,
\end{align}
where $\kappa^2 = R^{-3}\partial_R (R^4 \Omega^2)$ is the epicyclic frequency, $N^2_{R,z} = - (c_P \rho_\mathrm{g})^{-1}\partial_{R,z} P \partial_{R,z} S$ are the radial and vertical contributions to the buoyancy frequency, $S \equiv c_P \ln(P^{1/\gamma}\rho_\mathrm{g})$ is the gas specific entropy, $c_P$ is the specific heat capacity, and $\gamma$ is the adiabatic index. These criteria are satisfied in radially smooth and vertically thin discs \citep[][]{Lesur2023ASPC..534..465L}.

\par However, the above stability arguments assume that the gas behaves \emph{adiabatically} and that the cooling timescale is effectively infinite ($\tau_\mathrm{cool} \rightarrow \infty$). Thermo-hydrodynamical instabilities, such as the VSI, instead require a finite cooling time and a radial temperature gradient. This condition is satisfied in \emph{locally isothermal} discs ($\tau_\mathrm{cool} \rightarrow 0$) or in \emph{thermally relaxing} discs ($\tau_\mathrm{cool} > 0$) with temperature profiles of the form $T(R) \propto R^{-q}$, where $q \neq 0$. If the temperature $T$ follows a strict radial power law, the equilibrium angular velocity is explicitly a function of both radius and height, $\Omega(R,z)$. The disc therefore exhibits vertical differential rotation ($\partial_z \Omega \neq 0$). The VSI is thus a \emph{baroclinic} instability, since vertical shear arises only when the pressure depends on both density and temperature, $P(\rho,T)$, such that surfaces of constant density and pressure are misaligned $(\nabla \rho_\mathrm{g} \times \nabla P \neq 0)$ \citep[][]{Nelson2013MNRAS.435.2610N, Lesur2023ASPC..534..465L}.

\par The vertical shear provides the \emph{free energy} that drives the VSI. In rotating axisymmetric flows, this shear renders the disc susceptible to the Goldreich–Schubert–Fricke instability \citep{Goldreich1967ApJ...150..571G, Fricke1968ZA.....68..317F}. The instability develops when the stabilising effect of vertical buoyancy is weak or when the cooling time is sufficiently short ($\tau_\mathrm{cool}\lesssim\tau_\mathrm{dyn}$). Under these conditions, a displaced fluid element approximately conserves its initial angular momentum while rapidly adjusting its entropy to that of the surrounding gas \citep[][]{Nelson2013MNRAS.435.2610N, Lin2015ApJ...811...17L}.

\par For a locally isothermal disc, the growth rate of the axisymmetric ($k_y=0$) VSI modes following the definition of \citet{Latter2018MNRAS.474.3110L} is approximately
\begin{align}
    \sigma^2 \sim -\frac{k_z^2}{k^2}\Omega^2 \left( 1+2q_z \frac{k_x}{k_z} \right),\label{eq:growth_rates}
\end{align}
where the perturbation wavenumber is $k^2=k_x^2+k_z^2$ and $q_z=-R_0\partial \ln\Omega/\partial z$ is the vertical shear parameter. When $q_z\neq0$, unstable modes satisfy,
\begin{align}
    \frac{k_x}{k_z} < -\frac{1}{2q_z}. \label{eq:growing_modes}
\end{align}
In thin quasi-Keplerian discs ($H/R \ll 1$), the vertical shear is weak ($| R \partial_z \Omega| \sim \Omega_K H/R$, implying $q_z \ll 1$). Nevertheless, in the absence of buoyancy, the associated free energy can be released through modes that are radially short but vertically elongated ($k_x \ll k_z$) \citep[][]{Lesur2023ASPC..534..465L}.

\par The theory of the VSI in a radially unstratified framework has been developed for both the linear regime \citep[][]{Urpin1998MNRAS.294..399U, Nelson2013MNRAS.435.2610N, Barker2015MNRAS.450...21B, McNally2015ApJ...811..121M} and the weakly nonlinear regime \citep[][]{Latter2018MNRAS.474.3110L, Shtemler2019MNRAS.488.4207S, Shtemler2020MNRAS.499.3222S, Shtemler2025MNRAS.543.1313S}. The radial local VSI has been analysed analytically in the linear regime by \citet{Lehmann2023MNRAS.522.5892L}. To our knowledge, however, numerical simulations of the radially unstratified vertical shear instability have not yet been performed, although the nonlinear dynamics of the VSI have been extensively studied in quasi-\emph{global} disc models \citep[see e.g.][]{Nelson2013MNRAS.435.2610N, Stoll2014A&A...572A..77S, Flores2020A&A...644A..50F, Lesur2025A&A...703A.225L}.

\par If one moves from a radially global setup to a radially unstratified one, the radial power-law temperature profile that drives the classical VSI is no longer present \citep[][]{McNally2015ApJ...811..121M}. In this study, we therefore consider a \emph{barotropic} (globally isothermal) disc, in which the temperature structure does not produce vertical shear in the azimuthal velocity. Nevertheless, in a stratified shearing box \citep[][]{Goldreich1965MNRAS.130..125G}, a vertical dependence of the gas azimuthal velocity can still arise through dust-gas drag. In this case, the vertical shear is not thermally driven, as in the classical VSI, but dust-driven \citepalias{Lin2021ApJ...907...64L}.

\par This distinction is important because, in an isothermal shearing box without temperature gradients, the classical VSI is absent. The instability considered here is also distinct from the classical streaming instability (SI) \citep[see e.g.][]{Johansen2007ApJ...662..627J, Bai2010ApJ...722.1437B, Li2021ApJ...919..107L, Lim2025ApJ...993...12L}. In our setup, the simulations are initialised in hydrostatic equilibrium and include gas acceleration due to dust drag force, but neglect gas drag on the dust. The resulting instability is therefore not a drag instability \citep[][]{Squire2018MNRAS.477.5011S, Squire2018ApJ...856L..15S} in the same sense as the SI. Instead, the DVSI is more closely related to the vertically shearing streaming instability (VSSI; \citetalias{Lin2021ApJ...907...64L}), the stratified setup of \citet{Ishitsu2009arXiv0905.4404I}, and the protoplanetary-disc analogue of the baroclinic symmetric instability \citep[SymI;][hereafter \citetalias{Sengupta2023ApJ...942...74S}]{Sengupta2023ApJ...942...74S}. The main difference is that the DVSI studied here does not include dynamical dust drag with feedback but the dust drag on the gas is implemented as a fixed external force on the gas (the gas feels a time-invariant and height-dependent acceleration that represents the drag force on the gas in equilibrium).

\par We adopt this `dust-analogue’ approach, in which dust drag is represented as an external force on the gas, in order to isolate the DVSI linear modes from dust-drift instabilities like the SI. It is important to note, however, that this approximation breaks down once the perturbations reach sufficiently large amplitudes and the system enters the nonlinear, turbulent regime. In a fully coupled dust-gas system, the turbulence generated by the DVSI would stir and thicken the dust layer, thereby modifying the vertical shear profile that drives the instability. Neglecting this evolution is therefore likely to overestimate the forcing in the nonlinear state. The present work should consequently be regarded as an initial exploration of the VSSI, focusing specifically on its shear-driven component and the associated linear modes. A more complete treatment of the VSSI requires the inclusion of dynamical dust evolution and feedback, including the dust-drift contribution to the instability, as in the 2D axisymmetric simulations of \citet{Lehmann2025ApJ...983..135L}. These effects are investigated in the companion study \citet{MatthijsseInprepA&APaperII}, hereafter \citetalias{MatthijsseInprepA&APaperII}, where we examine its stabilising effect and present a detailed comparison between the DVSI and the VSSI.


\section{Equations of motion}\label{sec:eq_of_motions}

\par When studying the DVSI, we wish to isolate the vertical shear contribution. Therefore, we model the system using a single fluid, namely the gas. However, we still want the equilibrium profile to be dependent on the dust. We can do this by starting in hydrostatic equilibrium and solving for the Euler equations governing the linear and nonlinear evolution of gas:
\begin{align}
    \partial_t \rho_\mathrm{g} + \nabla \cdot (\rho_\mathrm{g} \mathbf{v}) &= 0, \label{eq:euler1}\\
    \partial_t \mathbf{v} + (\mathbf{v} \cdot \nabla) \, \mathbf{v} &= - \frac{\nabla p}{\rho_\mathrm{g}} + \mathbf{a}_\mathrm{g} + \mathbf{a}_{\mathrm{drag}, \, \mathrm{g}}, \label{eq:euler2}
\end{align}
where $p$ is the gas pressure, $\mathbf{a}_\mathrm{g}$ contains the forces acting on the gas other than dust drag, and $\mathbf{a}_{\mathrm{drag}, \, \mathrm{g}}$ denotes the acceleration associated with the dust back-reaction on the gas. Rather than evolving this term self-consistently with the dust, we compute it from the initial hydrostatic equilibrium and hold it fixed throughout the simulation. This preserves the dust-induced vertical shear while excluding the dynamical and stabilising response of the dust.

\par We simulate the DVSI in a radially unstratified, vertically stratified shearing box \citep{Goldreich1965MNRAS.130..125G}. The background quanties do not depend on the radial coordinates. The radially unstratified and vertically global VSI is similar to \citet{McNally2015ApJ...811..121M}, but their setup is adiabatic, while we consider an isothermal setup, where the vertical angular velocity profile $\mathbf{\Omega}(z)$ is not dependent on the internal energy density $e$ \citep[see,][]{McNally2015ApJ...811..121M}, but the vertical dependency of the drag force $\mathbf{a}_{\mathrm{drag}, \, \mathrm{g}}$. 

\par Setting up a stratified shearing-box will introduce inertial forces: the Coriolis force, tidal force, and to make the box stratified, a vertical component of stellar gravity is added. These forces are included through an effective potential $\Phi = -S\Omega x^2 + \Omega^2z^2/2$, where $\Omega$ is the angular velocity and $S$ the shearing rate, given by $S = \frac{3}{2}\Omega$ in a Keplerian disc. The global pressure gradient ($\partial P/\partial R$) is parameterized with\footnotemark
\begin{align}
    \eta = \frac{1}{2\rho_\mathrm{g}} \frac{\partial P}{\partial R},
    \label{eq:pressure_gradient}
\end{align}
in the local setup.
\footnotetext{We use a dimensional pressure support parameter $\eta$ similar to \citet{Paardekooper2020MNRAS.499.4223P, Paardekooper2021MNRAS.502.1579P, McNally2021MNRAS.502.1469M}. This parameter is related to the non-dimensional parameter defined in \citet{Youdin2005ApJ...620..459Y} by $\eta = r_0 \Omega^2 \eta_\mathrm{YG}$. This is a largely cosmetic choice}

\par Rewriting equations \eqref{eq:euler1} and \eqref{eq:euler2}, with the effective potential and forces in the radially unstratified, vertically stratified setup, gives us,
\begin{align}
    \partial_t\rho_\mathrm{g}+\nabla \cdot (\rho_\mathrm{g} \mathbf{v}) =& \, 0,\label{eq:euler1_local}\\
    \partial_t \mathbf{v} + (\mathbf{v} \cdot \nabla) \mathbf{v} =& \, 2\eta\mathbf{\hat{x}}-\frac{1}{\rho_\mathrm{z}}\frac{\nabla p}
    {\rho_\mathrm{g}}-2\mathbf{\Omega}\times \mathbf{v} - \nabla\Phi \nonumber + \mathbf{a}_\mathrm{drag, \, g} \\
    &+ \frac{\nabla \cdot \mathsf{T}}{\rho_\mathrm{g}}.\label{eq:euler2_local}
\end{align}
Here, we add gas viscosity through the viscous stress tensor:
\begin{align}
    \mathsf{T} = \rho_\mathrm{g} \nu \left[ \nabla \mathbf{v} + (\nabla \mathbf{v})^T -\frac{2}{3} \mathsf{I} \nabla \cdot \mathbf{v} \right],\label{eq:viscous_tensor}
\end{align}
where $\nu$ is the kinematic viscosity, which is implemented with the $\alpha$-prescription
\begin{align}
    \nu = \frac{\alpha_\mathrm{g} c_\mathrm{s}^2}{\Omega}\frac{\rho_\mathrm{g, \, eq}(z)}{\rho_\mathrm{g}(z)},\label{eq:viscosiity}
\end{align}
with the dimensionless gas diffusion coefficient $\alpha_\mathrm{g}$, quantifying the strength of unresolved turbulence \citep{Shakura1973A&A....24..337S}. The gas density fraction ensures that the viscous overstability in an isothermal disc is eliminated and $\rho_\mathrm{g}\nu$ is constant in the equilibrium state when $\rho_\mathrm{g}(z) = \rho_\mathrm{g, \, eq}(z)$ \citep{Latter2006MNRAS.372.1829L, Chen2020ApJ...891..132C}.


\subsection{Equilibrium state}\label{sec:equilibrium}

\begin{figure}[ht!]
    \centering
    \includegraphics[width=\hsize]{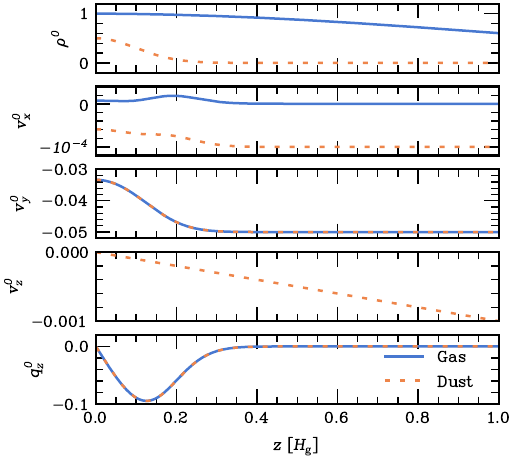}
    \caption{Two-fluid equilibrium profile for standard parameters ($\alpha_D = 10^{-5}; \, Z = 0.05; \, \eta = 0.05; \, \tau_\mathrm{s} = 10^{-3}$), showing the densities and velocities for the dust and gas, and the vertical shear of the azimuthal velocity $q_z$ for the gas and dust. From initial assumption $v_{z}^0=0$.}\label{fig:equilibrium_profiles}
\end{figure}

In \citetalias{Lin2021ApJ...907...64L}, vertical equilibrium profiles were calculated for a dust-gas mixture. These depend on the Stokes number $\tau_\mathrm{s}$ and the dust diffusion coefficient\footnote{The implementation of the dust diffusion in \texttt{FARGO3D} does not affect the dust’s vertical velocity $u_{z}$, which is set by the vertical gravity $u_{z}=-\tau_\mathrm{s} \Omega$. Dust diffusion depends only on the dust density gradient. The mass flux from the vertical diffusion is balanced by the dust settling velocity.} $\alpha_D$. From this solution, we calculate $\mathbf{a}_{\mathrm{drag}, \, \mathrm{g}}$ to be inserted in equation \eqref{eq:euler2}. The equilibrium gas profile then automatically matches that of Fig.~\ref{fig:equilibrium_profiles}.

\par The vertical structure of the densities and vertical velocities can be obtained analytically, whereas the horizontal velocities must be computed numerically. These velocities are obtained by solving a boundary value problem on an infinite interval. The source code used to compute this two-fluid equilibrium is publicly available in two GitHub repositories\footnote{\url{https://github.com/SijmeJan/psi-stratified}\\ \url{https://github.com/SijmeJan/spectral-solvers}}. With this code we obtain the gas density $\rho_\mathrm{g}^0$, the gas velocities $v_{x}^0$, $v_{y}^0$ (excluding the background shear), and $v_{z}^0$, as well as the dust density $\rho_\mathrm{d}^0$ and velocities $\mathbf{u}^0$. These quantities depend on the dust diffusion coefficient $\alpha_D$, the metallicity $Z$, the background pressure gradient $\eta$, the Stokes number $\tau_\mathrm{s}$, and the vertical coordinate $z$. The resulting profiles are used as the initial conditions for the \texttt{FARGO3D} simulations.  An example of the resulting density and velocity profiles for gas and dust, together with the shear profile\footnote{This differs from the shear profile of \citet{Latter2018MNRAS.474.3110L}.} ($q_z^0=\partial_z v_y^0$), is shown in Fig.~\ref{fig:equilibrium_profiles} (see also Fig.~1 of \citetalias{Lin2021ApJ...907...64L}).

\par We use dimensionless units by choosing a timescale $\Omega^{-1}$ and a length scale of the gas scale height $H_\mathrm{g}$ (the units are $\Omega \! = \! H_\mathrm{g} \! = \! c_\mathrm{s} \! = \! 1$). When the timescale is $\Omega^{-1}$, the parameters governing the system are now the background pressure gradient $\eta$ and has a default value of $\eta \! = \! 0.05$, the shear parameter $S/\Omega$, the dust-to-gas ratio $\mu= \rho_\mathrm{d}/\rho_\mathrm{g}$, the metallicity $Z$, and the dust scale height $H_\mathrm{d}$. The dust scale height is dependent on the stirring due to the vertical component of turbulence $\alpha_{D}$ and the Stokes number $\tau_\mathrm{s}$, which can be approximated by,
\begin{align}
    H_\mathrm{d} = \sqrt{\frac{\alpha_D}{\tau_\mathrm{s}}}H_\mathrm{g},\label{eq:scale_height}
\end{align}
\citep{Dubrulle1995Icar..114..237D,Youdin2007Icar..192..588Y}. The metallicity is defined by the ratio between the dust and gas surface density,
\begin{align}
    Z \equiv \frac{\Sigma_\mathrm{d}}{\Sigma_\mathrm{g}} = \frac{\int_{-\infty}^{\infty}\rho_\mathrm{d}\mathrm{d}z}{\int_{-\infty}^{\infty}\rho_\mathrm{g}\mathrm{d}z} \simeq \mu_\mathrm{z=0} \frac{H_\mathrm{d}}{H_\mathrm{g}}.\label{eq:metallicity}
\end{align}


\section{Numerical approach}\label{sec:numerical_aprouch}

\par We evaluate the initial growth and the saturated nonlinear regime of the DVSI using numerical simulations with the hydrodynamical code \texttt{FARGO3D} \citep{Benitez-Llambay2016ApJS..223...11B, Benitez-Llambay2019ApJS..241...25B}. We want to start close to equilibrium to prevent spurious perturbations from growing while the system settles towards the equilibrium velocities. Therefore, the simulations are initialised in hydrostatic equilibrium using the two-fluid solutions of \citetalias{Lin2021ApJ...907...64L}. On top of the equilibrium velocities, we add white noise to seed the instability. The analytical solutions of horizontal velocity profiles from the two-fluid approach of \citetalias{Lin2021ApJ...907...64L}, need to be solved numerically.

\par To implement the dust-analogue, we add a \emph{fixed} drag acceleration $\mathbf{a}_\mathrm{drag,g}$ to the \texttt{FARGO3D} source code. This acceleration depends only on the equilibrium densities and velocities at $\Omega t=0$. Consequently, the drag acceleration on the gas remains constant as the instability grows and does not depend on the instantaneous velocity difference between gas and dust. Using this “dust-analogue” approach, the equilibrium state remains consistent with the two-fluid continuity and momentum equations of \citetalias{Lin2021ApJ...907...64L} (see Sect.~\ref{sec:equilibrium}).

\par The dust analogues assume a constant Stokes number and that the fluid approximation applies ($\tau_\mathrm{s} \ll 1$). A constant Stokes number does not correspond to a constant particle size in a stratified disc, because the gas density varies with height and the Stokes number depends on the gas density. As a result, fixing $\tau_\mathrm{s}$ throughout the simulation box implies that the effective particle size represented by the dust fluid increases with height.


\subsection{The tracer fluid}\label{sec:tracer_fluid}

\par In all \texttt{FARGO3D} simulations we include two additional tracer fluids that do not influence the gas dynamics; the gas does not experience their drag force. However, the tracer fluids do respond to the gas velocity field and vertical gravity, allowing them to trace the turbulent mixing of the gas.

\par The tracers are strongly coupled to the gas with Stokes numbers $\tau_\mathrm{s}=10^{-2}$ and $\tau_\mathrm{s}=10^{-6}$. They are initially confined to a thin layer around the midplane with scale height $H_\mathrm{tracer}/H_\mathrm{g}=10^{-2}$ and have an initial surface density fraction $\Sigma_\mathrm{tracer}/\Sigma_\mathrm{g}=10^{-2}$. No diffusion is applied to the tracer fluids ($\alpha_\mathrm{tracer}=0$). As the DVSI grows and saturates, the tracers are therefore dispersed only by gas turbulence. The Stokes number of $\tau_\mathrm{s}=10^{-2}$ is used as a tracer to calculate the dust diffusion coefficient $\alpha_D$ using Eq.~\eqref{eq:scale_height}, whilst the extremely tightly coupled tracer with Stokes number of $\tau_\mathrm{s}=10^{-6}$ is used to track the movement of the gas.


\subsection{Boundary conditions}\label{sec:boundary}

\par In the stratified shearing box, we use periodic boundary conditions in the azimuthal direction and shear-periodic boundaries in the radial direction. In the vertical direction, the boundaries must remain consistent with hydrostatic equilibrium. The densities and velocities in ghost cells are therefore given by,
\begin{align}
    \rho_\mathrm{gh} =& \, \left(\rho_\mathrm{act}-\rho_\mathrm{act}^0 \right) \left( 1-\frac{\Delta t}{\tau_\mathrm{damp}}\right)+\rho_\mathrm{gh}^0,\\
    v_\mathrm{gh} =& \, \left( v_\mathrm{act}-v_\mathrm{act}^0 \right)\left( \frac{\rho_\mathrm{act}}{\rho_\mathrm{gh}} \right)\left( 1-\frac{\mathrm{dt}}{\tau_\mathrm{damp}}\right)+v_\mathrm{gh}^0.
\end{align}
With this prescription, the background equilibrium $f^0$ is continued into the ghost cells, while the perturbation from the adjacent active cell is added. The boundary condition is symmetric with respect to the perturbation ($ f’_\mathrm {act} = f’_\mathrm {gh}$), where $f = f^0 + f'$. Both inflow and outflow are allowed, but perturbations are damped with a timescale $\tau_\mathrm{damp}$ to suppress boundary noise during inflow. In this work we adopt $\tau_\mathrm{damp} = 10^{-3}$.

\par In addition to the damping applied in the ghost cells, we include gas diffusion near the vertical boundaries through a vertically dependent $\alpha$-diffusion,
\begin{align}
    \alpha(z) =  8.5\cdot10^{-18} \cosh{\left( \frac{40z}{z_\mathrm{max}}\right)} \, \alpha_\mathrm{boundary},
\end{align}
The numerical factors $8.5\times10^{-18}$ and $40$ are chosen such that the diffusion coefficient reaches $\alpha_\mathrm{boundary}$ at the vertical boundary ($\alpha(z_\mathrm{max}) \approx \alpha_\mathrm{boundary}$). This profile damps strong gradients near the boundary, reducing wave reflections while ensuring that diffusion does not influence the region where the DVSI develops. In this work we adopt $\alpha_\mathrm{boundary} = 10^{-4}$.

 
\begin{table}[ht!]
    \caption{List of simulation runs.}     
    \label{tab:simulations}      
    \centering 
    \setlength{\tabcolsep}{4pt}
    \renewcommand{\arraystretch}{1.4}
    \begin{tabular}{l c c c c c}     
        \hline\hline       
        \textbf{Run Name} & $\mathbf{N}_\mathrm{grid}$ & $\mathbf{Z}$ &  $\boldsymbol{\alpha}_\mathbf{D}$ &  $\boldsymbol{\tau}_\mathbf{s}$ & $\boldsymbol{\alpha}_\mathbf{g}$  \\
        \hline
        \texttt{DVSI} & $1024$ & $0.05$ & $10^{-5}$ & $10^{-3}$ & $0$ \\
        \hline
        \texttt{DVSI}$_{N,\, 256}$ & $512$ & $0.05$ & $10^{-5}$ & $10^{-3}$ & $0$ \\
        \texttt{DVSI}$_{N,\, 512}$ & $512$ & $0.05$ & $10^{-5}$ & $10^{-3}$ & $0$ \\
        \texttt{DVSI*}$_{N,\, 1024}$ & $1024$ & $0.05$ & $10^{-5}$ & $10^{-3}$ & $0$ \\
        \texttt{DVSI*}$_{N,\, 2048}$ & $2048$ & $0.05$ & $10^{-5}$ & $10^{-3}$ & $0$ \\
        \hline
        \texttt{DVSI}$_{Z,\, 0.01}$ & $1024$ & $0.01$ & $10^{-5}$ & $10^{-3}$ & $0$ \\
        \texttt{DVSI}$_{Z,\, 0.025}$ & $1024$ & $0.025$ & $10^{-5}$ & $10^{-3}$ & $0$ \\
        \texttt{DVSI}$_{Z,\, 0.1}$ & $1024$ & $0.1$ & $10^{-5}$ & $10^{-3}$ & $0$ \\
        \texttt{DVSI}$_{Z,\, 0.5}$ & $1024$ & $0.5$ & $10^{-5}$ & $10^{-3}$ & $0$ \\
        \hline
        \texttt{DVSI}$_{\alpha_D,\, 1\mathrm{e}-6}$ & $1024$ & $0.05$ & $10^{-6}$ & $10^{-3}$ & $0$ \\
        \texttt{DVSI}$_{\alpha_D,\, 1\mathrm{e}-4}$ & $1024$ & $0.05$ & $10^{-4}$ & $10^{-3}$ & $0$ \\
        \hline
        \texttt{DVSI}$_{\tau_\mathrm{s}, 1\mathrm{e}-4}$ & $1024$ & $0.05$ & $10^{-5}$ & $10^{-4}$ & $0$ \\
        \texttt{DVSI}$_{\tau_\mathrm{s}, 1\mathrm{e}-2}$ & $1024$ & $0.05$ & $10^{-5}$ & $10^{-2}$ & $0$ \\
        \hline
        \texttt{DVSI}$_{\alpha_\mathrm{g},\, 1\mathrm{e}-7}$ & $1024$ & $0.05$ & $10^{-5}$ & $10^{-3}$ & $10^{-7}$ \\
        \texttt{DVSI}$_{\alpha_\mathrm{g},\, 1\mathrm{e}-6}$ & $1024$ & $0.05$ & $10^{-5}$ & $10^{-3}$ & $10^{-6}$ \\
        \texttt{DVSI}$_{\alpha_\mathrm{g},\, 1\mathrm{e}-5}$ & $1024$ & $0.05$ & $10^{-5}$ & $10^{-3}$ & $10^{-5}$ \\
        \hline\hline
        \multicolumn{2}{c}{$\left[\mathbf{L_x \times L_z}\right]$}& \multicolumn{1}{c}{$\boldsymbol{\eta}$} & \multicolumn{2}{c}{$\boldsymbol{\rho}^0_\mathrm{g,\, mid}$} & \multicolumn{1}{c}{$\mathbf{\hat{X}}$} \\
        \multicolumn{2}{c}{$\left[ 1\times8 \right] H_\mathrm{g}$; if \texttt{*} $L/2$} & \multicolumn{1}{c}{$0.05$} & \multicolumn{2}{c}{$1$} & \multicolumn{1}{c}{$10^{-3}$} \\
        \hline
    \end{tabular}
    \tablefoot{
    Simulation parameters, with $N_\mathrm{grid}$ the spatial resolution in units of number of cells per gas scale height, $Z$ the metallicity, $\alpha_D$ and $\alpha_\mathrm{g}$ the dust and gas diffusion coefficient respectifly, $\tau_\mathrm{s}$ the Stokes number, $L$ the size of the shearing box (which is halfed for the two runs with a \{\texttt{*}\}), $\eta$ the background pressure gradient, $\rho^0_\mathrm{g, mid}$ the gas density at the mid plane, and $\hat{X}$ the amplitude of the white noise perturbation.
    }
\end{table}

\subsection{Setup}

\par The DVSI is simulated in a two-dimensional shearing box spanning the radial ($\hat{\mathbf{x}}$) and vertical ($\hat{\mathbf{z}}$) directions. The simulations are initialised with the equilibrium density and velocity profiles obtained from the solutions of the momentum equations from \citetalias{Lin2021ApJ...907...64L}, (see Fig.~\ref{fig:equilibrium_profiles}). On top of the equilibrium velocities, white noise is added with an amplitude $\hat{X}=10^{-3}c_\mathrm{s}$.

\par These equilibrium profiles depend on the dust diffusion coefficient $\alpha_D$, the metallicity $Z$, the background pressure gradient $\eta$, and the Stokes number $\tau_\mathrm{s}$. Together with the box size $L_{x,z}$, spatial resolution $N_\mathrm{grid}$, and the gas diffusion coefficient $\alpha_\mathrm{g}$, these parameters define the simulation setup.

\par The shearing box used in our simulations is elongated in the vertical direction because the fastest-growing DVSI modes have much smaller radial than vertical wavelengths (see Eq.~\eqref{eq:growing_modes}). In this study we adopt a box size of $L_x=1H_\mathrm{g}$ and $L_z=8H_\mathrm{g}$ with a spatial resolution of $N_\mathrm{grid}=1024/H_\mathrm{g}$. The gas density at the midplane is normalised to $\rho^0_{\mathrm{g,mid}}=1$. These parameters, together with the others described above, are summarised in Table~\ref{tab:simulations}.


\section{Dust Driven Vertical Shear Instability}\label{sec:DVSI}

\par The DVSI is initialised by the equilibrium solutions with an imposed random noise on top of the velocity profiles, such that all modes are excited, and growth is expected wherever vertical shear is present ($q_z \neq 0$). For the DVSI, this occurs throughout the domain except at the midplane ($z=0$; see Fig.~\ref{fig:equilibrium_profiles}). The growth rate of the perturbations depends on the strength of the vertical shear and on the orientation of the modes, characterised by $k_x/k_z$ (Eq.~\eqref{eq:growth_rates}). When the shear is driven by dust drag force, the regions of strongest vertical shear correspond to locations where the vertical gradient of the dust-to-gas ratio, $\partial_z \mu(z)$, is largest. The growth rate depends only on the wave orientation and the background shear (see Eq.~\eqref{eq:growth_rates}). The background shear does vary with height $z$.


\begin{figure}[ht!]
    \includegraphics[width=\hsize]{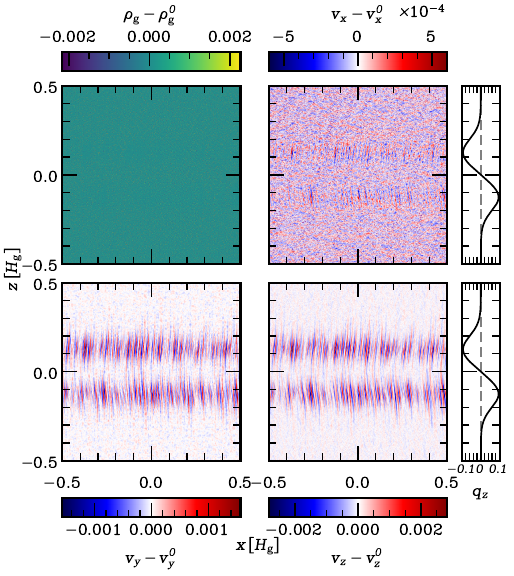}
    \caption{Heatmap of run \texttt{DVSI} at $\Omega t=50$, showing the perturbed densities and velocities in all directions, where the background densities and velocities are subtracted. The sidebar shows the azimuthal component of the vertical shear profile. The heatmaps do not cover the full vertical extent of the simulation of $z_\mathrm{max}=4H_\mathrm{g}$, only showing $L_z/8$.}
    \label{fig:DVSI_heatmap_gas_t50}
\end{figure}

\subsection{Linear growth}

\par Because the vertical shear varies with height, the amplitude and orientation of the perturbation modes are maximal at specific vertical locations. Figure~\ref{fig:DVSI_heatmap_gas_t50} shows the perturbations in density and velocity for run \texttt{DVSI} at $\Omega t = 50$, prior to the onset of the nonlinear regime. The gas (top right panel), still shows some white noise on the grid cell level from the initial white noise seed, which was added on top of the equilibrium velocities. In the azimuthal and vertical velocity components ($v_y$ and $v_z$), the height dependence of the growth rate is visible as two pronounced bands located at the vertical positions where $|q_z|$ is largest. The dependence of the mode orientation ($k_x/k_z$) on $q_z$ is also apparent in the S-shaped structure of the velocity perturbations, where the tilt of the modes is strongest near the heights at which $q_z$ peaks.

\par Figure~\ref{fig:DVSI_heatmap_gas_t50} also shows that $k_x \gg k_z$, because the perturbation modes are nearly vertical. This is further reflected in the relative amplitudes of the velocity perturbations: the vertical component reaches much larger values ($\max(| v’_z |) \gg \max(| v’_x |)$), while the radial perturbations remain close to the background noise level. The radial wavenumber is initially large and decreases with time, indicating that the fastest-growing modes correspond to the highest wavenumbers. These linear modes are also visible in Fig.~3 of \citet{Lehmann2025ApJ...983..135L} where 2D axisymmetric simulations show signatures of the VSSI, because the work of \citet{Lehmann2025ApJ...983..135L} does include dynamical dust feedback. This suggests that when the VSSI is shear dominated it behave similar to the DVSI in the linear regime.


\begin{figure}[ht!]
    \centering
    \includegraphics[width=\hsize]{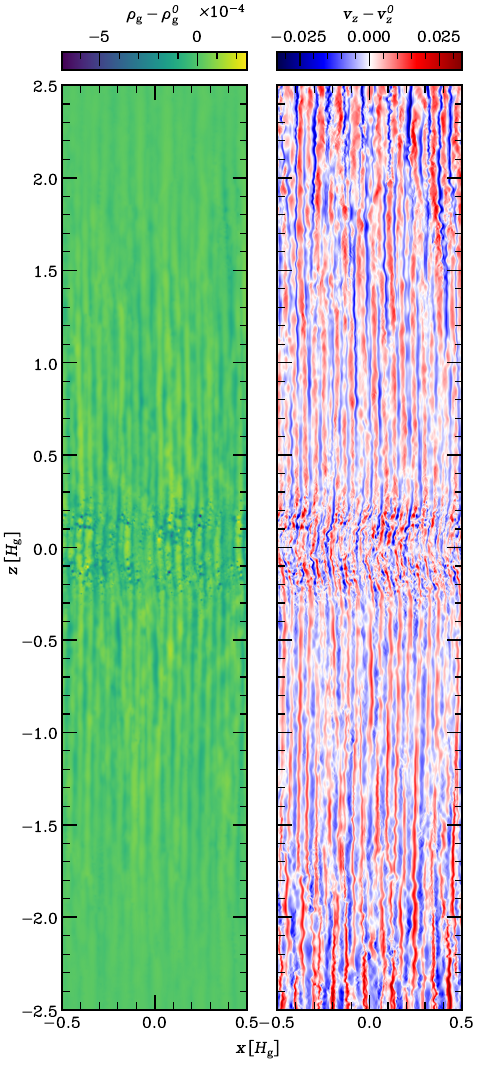}
      \caption{Heatmap of run \texttt{DVSI} at $\Omega t = 500$, showing the perturbed densities and vertical velocity, where the background densities and velocities are subtracted. The heatmaps do not cover the full vertical extent of the simulation of $z_\mathrm{max}=4H_\mathrm{g}$, only showing $5/8L_z$.}
    \label{fig:DVSI_heatmap_gas_t500}
\end{figure}

\subsection{Non-linear saturation}

\par The perturbations are largest in the regions of strong vertical shear, but the resulting velocity structures extend well beyond these regions and cross the midplane, where $q_z = 0$. As a result, the mixing motions extend throughout the vertical domain of the simulation box. This is illustrated in Fig.~\ref{fig:DVSI_heatmap_gas_t500}, where the vertical extent of the modes reaches the full simulated domain up to $z = 4$ (Fig.~\ref{fig:DVSI_heatmap_gas_t500} is only showing up to $z=2.5$). That at higher heights the velocities increase further could be explained by the decreasing background density: in order to maintain continuous momentum transport, the velocity amplitudes must increase.

\par The DVSI develops into inertial waves where $\mathbf{k_x \gg k_z}$, with strong vertical velocities, and the vertical velocity of adjacent wave crest alternates. These modes are corrugation modes and are also observed in the pure hydro version of the “classical" VSI \citep[see e.g.][]{Nelson2013MNRAS.435.2610N}. The difference in direction of the vertical velocity between neighbouring modes causes strong radial shear of the vertical velocities that can saturate the DVSI.  This shear between the neighbouring modes drives a parasitic instability, similar to that described in Sect.~3 of \citet{Latter2018MNRAS.474.3110L}. Once the amplitudes become sufficiently large, the radial shear in the vertical velocity will develop, a Kelvin–Helmholtz-like instability, hereafter KH-like instability. In the present axisymmetric configuration, the DVSI produces extended modes in which vortices form in the radial–vertical ($x,z$) plane, rather than in the ($y,z$) plane, which differs from the classical picture of the Kelvin–Helmholtz instability, hereafter KHI. This parasitic instability limits the DVSI growth via the transfer of energy to smaller scales, generating an effective turbulent viscosity which can set their characteristic amplitude \citep[see Sect.~3.3.2][]{Latter2018MNRAS.474.3110L}.

\par
\begin{figure}[ht!]
    \includegraphics[width=\hsize]{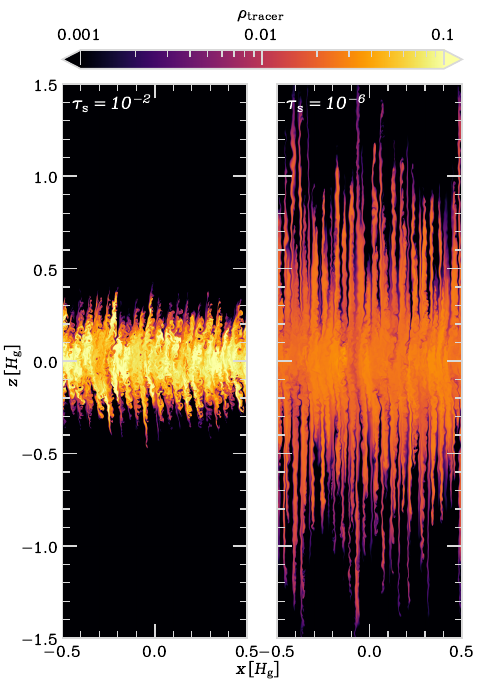}
    \caption{Heatmap of run \texttt{DVSI} at $\Omega t=500$, showing the perturbed densities of the tracer fluids with $\tau_\mathrm{s} = 10^{-2}$ (left) and $\tau_\mathrm{s} = 10^{-6}$ (right). The heatmaps do not cover the full vertical extent of the simulation of $z_\mathrm{max}=4H_\mathrm{g}$, only showing $3/8L_z$.}
    \label{fig:DVSI_heatmap_tracer_t500}
\end{figure}

 The secondary KH-like instability produces the small-scale eddies superimposed on the extended DVSI modes \citep[similar to][]{MelonFuksman2024A&A...682A.140M}. Some of these eddies form closed vortices in the radial–vertical ($x,z$) plane. Their presence is visible in Fig.~\ref{fig:DVSI_heatmap_gas_t500} in the perturbed density field, where they appear as small dots corresponding to pressure minima at the centres of the vortices, and in Fig.~\ref{fig:DVSI_heatmap_tracer_t500}, where the vortices are traced by the tracer fluids, appearing as whirls. In the radial velocity perturbations, the characteristic radial length scale is set by $k_x$, while their vertical extent remains limited to the characteristic scale of the vortices. By contrast, the vertical velocity perturbations retain the vertically extended, low-$k_z$ structure characteristic of the DVSI, although these modes are partially disrupted by the eddies. In the saturated state, the effect of the vortices becomes comparable to the DVSI near the midplane, which could explain why the maximum velocities in the radial and vertical directions become similar. The vortices are approximately isotropic, unlike the extended DVSI modes at higher $z$ values (see Sect.~\ref{sec:reynolds_stress}). In other parameter regimes, the saturated state is vortex-dominated rather than characterised by extended vertical modes (see Sect.~\ref{sec:parameterstudy}).


\subsection{Vertical mixing}\label{sec:vertical_mixning}

\par In our \texttt{FARGO3D} simulations, we use three fluids: the gas and two tracer fluids that follow the gas dynamics. The tracer fluids do not affect the gas, but they are accelerated by it through aerodynamic coupling, which depends on the Stokes number. They are also subject to the vertical component of gravity, but tracer fluids have no added sub-resolution diffusion to counteract settling. The tracer fluids are stirred by the gas turbulence, allowing the gas turbulence strength to be estimated from the relation between the tracer scale height and the turbulent diffusion coefficient (Eq.~\eqref{eq:scale_height}). Averaged over $300 \leq \Omega t \leq 750$, the tracer with $\tau_\mathrm{s} = 10^{-2}$ reaches a scale height of $H_\mathrm{tracer} = 0.138(6)H_\mathrm{g}$, corresponding to a turbulence level of $\alpha_\mathrm{g} = 1.91(11) \times 10^{-4}$, assuming a Schmidt number $\mathrm{Sc} \equiv \alpha_\mathrm{g}/\alpha_\mathrm{D} = 1$ \citep{Cuzzi1993Icar..106..102C, Youdin2007Icar..192..588Y}.

\par Because the tracers are not initialised in vertical equilibrium, but instead start in a thin layer around the midplane, they can also be used to probe the spatial inhomogeneity of the turbulence. Figure~\ref{fig:DVSI_heatmap_tracer_t500} shows the density distribution of the two tracer fluids for run \texttt{DVSI} at $\Omega t = 500$. The left panel corresponds to the more weakly coupled tracer ($\tau_\mathrm{s}=10^{-2}$), which has already reached its vertical equilibrium but still exhibits strong vertical fluctuations in the form of spikes (vertically extended thin columns in the tracer’s density).

\par These spikes arise from the vertical DVSI modes, but they do not persist indefinitely: each spike grows to a certain height before collapsing again. This behaviour can be explained by the oscillatory motion of the extended vertical modes of the DVSI, in which the vertical velocity periodically reverses direction. When a spike collapses, new spikes develop at neighbouring radii, where another vertical channel switches its vertical velocity component.

\par
\begin{figure}[ht!]
    \includegraphics[width=\hsize]{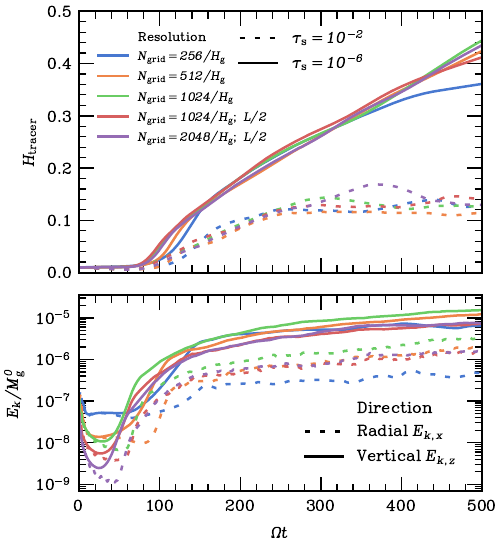}
    \caption{Time evolution of the tracer fluids scale heights (top) and directional kinetic energy normalized by the inital gas mass (bottom) for runs \texttt{DVSI}$_{N, \, 256}$, \texttt{DVSI}$_{N,\, 512}$, \texttt{DVSI}, \texttt{DVSI*}$_{N,\, 1024}$ and \texttt{DVSI*}$_{N,\, 2048}$.}
    \label{fig:DVSI_Ngrid_Htracer_Ek}
\end{figure}

 The maximum height and turnover of these spikes allow the radially averaged density distribution to reach equilibrium once the extended vertical modes stop growing and radial diffusion produced by the secondary instability smooths the radial density variations, making the spikes less pronounced. This behaviour is visible in the time evolution of the tracer scale height shown in the top panel of Fig.~\ref{fig:DVSI_Ngrid_Htracer_Ek}, where the green dotted line corresponds to the $\tau_\mathrm{s}=10^{-2}$ tracer of run \texttt{DVSI}. The tracer reaches equilibrium after approximately $250 \, \Omega t$. The scale height is calculated by 
\begin{align}
    H_\mathrm{tracer}= \sqrt{\frac{\int z^2\rho_\mathrm{tracer} \mathrm{d}z}{\int\rho_\mathrm{tracer}\mathrm{d}z}},\label{eq:tracer_scale_height}
\end{align}
at each snapshot.

\par The more strongly coupled tracer (right panel of Fig.~\ref{fig:DVSI_heatmap_tracer_t500}) has not yet reached equilibrium (see solid green line in top panel of Fig.~\ref{fig:DVSI_Ngrid_Htracer_Ek}). In this case, new spikes overtake the locations of older ones, reaching larger heights because the extended modes are still growing, and tracer is less affected by radial diffusion at these heights because the turbulence is a lot more anisotropic further from the midplane.


\subsection{Convergence} \label{sec:convergence}

\par In the linear regime of the DVSI, all wave numbers are unstable wherever $q_z \neq 0$ and Eq.~\eqref{eq:growing_modes} is satisfied. This implies that progressively larger wavenumbers are also unstable and can grow (even faster than smaller wavenumbers). However, in numerical simulations, the largest resolvable wavenumber depends on the grid resolution $N_\mathrm{grid}$. In the linear regime (Fig.~\ref{fig:DVSI_heatmap_gas_t50}), the fastest-growing modes are therefore determined by the highest radial wavenumbers that can be resolved. We therefore do a convergence study where we investigate the growth by evaluating the time evolution of the tracer scale height and the directional kinetic energy at different grid resolutions and box sizes, the time evolutions are shown in Fig.~\ref{fig:DVSI_Ngrid_Htracer_Ek}. At the two lower resolutions, $N_\mathrm{grid} = 256$ and $512$ (blue and orange, respectively), the faster-growing modes that where resolved in higher resolution runs are not resolved, and the instability therefore develops more slowly than in the standard-resolution run $N_\mathrm{grid}=1024$ (green and red). 

\par In the saturated regime, the effect of resolution is much weaker. The saturated tracer scale height for $\tau_\mathrm{s} = 10^{-2}$ and the directional kinetic energy converge to similar values by $\Omega t = 500$, although run \texttt{DVSI}$_{N,\,256}$ remains a mild outlier. This behaviour can be visualised in the heatmaps shown in Fig.~\ref{fig:DVSI_Ngrid_t500}, where the dominant $k_x$ mode is well resolved and appears at a similar spatial scale across the different resolutions. 

\par The largest difference between the runs is the size of the eddies produced by the secondary parasitic KH-like instability. These eddies do not exhibit a single dominant wavenumber but instead form a turbulent cascade. The corresponding kinetic energy power spectrum (not shown) follows a power law, with a slope that shows behaviour similar to that reported by \citet{MelonFuksman2024A&A...682A.140M} and \citet{Lesur2025A&A...703A.225L}. Higher-resolution simulations progressively resolve smaller eddies. In run \texttt{DVSI}${N,\,256}$, the grid scale is close to the dominant $k_x$ scale of the extended vertical modes, and the secondary instability is therefore under-resolved. As a result, the low-resolution run behaves similarly to simulations that include explicit gas diffusion, such as run \texttt{DVSI}${\alpha_\mathrm{g},\,1\mathrm{e}-6}$ (see Sect.~\ref{sec:gas_diffusion}).

\par We also consider simulations with a smaller box size, $[L_x \times L_z] = [0.5 \times 4]H_\mathrm{g}$, at the same resolution as the standard run and a higher resolution, corresponding to $N_\mathrm{grid}=1024/H_\mathrm{g}$ and $2048/H_\mathrm{g}$ (runs \texttt{DVSI*}$_{N,\,1024}$ and \texttt{DVSI*}$_{N,\,2048}$). These simulations exhibit similar growth rates and saturate at the same levels as run \texttt{DVSI}. This demonstrates that the numerical domain is larger than the characteristic scale of the instability, and that a resolution of $N_\mathrm{grid}=1024/H_\mathrm{g}$ is sufficient for convergence.


\begin{figure}[ht!]
    \includegraphics[width=\hsize]{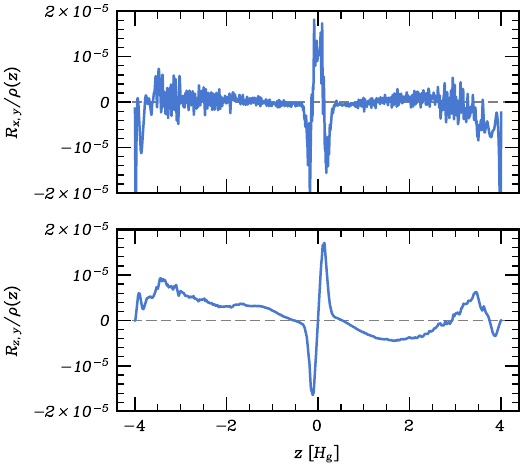}
    \caption{The specific stress tensor in the radial (top) and vertical (bottom) direction of run \texttt{DVSI}. The stress tensor is calculated from the radial average of the full simulation box and time-averaged over $\mathbf{500} \leq \Omega t \leq \mathbf{750}$.}
    \label{fig:DVSI_reynolds}
\end{figure}

\subsection{Reynolds stress tensor}\label{sec:reynolds_stress}

\par Similar to the classical VSI, DVSI-driven turbulence is anisotropic and vertically structured. We quantify the turbulent transport using the Reynolds stress components following \citet{Stoll2017A&A...599L...6S} and \citet{Krapp2020MNRAS.497.2715K}. The height-dependent stress is computed as
\begin{align}
    R_{i,y}(z) = \left\langle \rho_\mathrm{g} v_i \, \delta v_y \right\rangle_{x,t} \ ,
\end{align}
where $i$ denotes either the radial or vertical component, $v_i$ is the corresponding gas velocity, $\delta v_y$ is the perturbation of the azimuthal gas velocity relative to the background shear ($\delta v_y = v_{y} - v^0_{y} + Sx$), and the brackets denote an average over radius and time. The resulting stress profiles are shown in Fig.~\ref{fig:DVSI_reynolds}.

\par The turbulence is strongest within approximately one dust scale height, where both the radial and vertical stress components correspond to\footnote{This diffusion factor is $\sim\mathcal{O}(1)$ smaller than the diffusion measured with the tracer fluid in Sect.~\ref{sec:vertical_mixning}.} $\alpha \! \sim \! 10^{-5}$. In this region, the turbulent transport is approximately isotropic. At larger heights, the radial component becomes negligible, while vertical transport remains present. The turbulence, therefore, becomes anisotropic away from the dust layer, although the degree of anisotropy is substantially weaker than in “classical” VSI simulations. In particular, we do not find vertical transport exceeding radial transport by factors of several hundred, as reported by \citet{Stoll2017A&A...599L...6S}.

\par The Reynolds-stress-based diffusion coefficient is comparatively smaller than the effective $\alpha$ inferred from the tracer scale height. This difference can be explained by the spatial scale of the corrugation modes, which extend coherently across the full computational domain and are therefore efficient at displacing the entire dust midplane vertically. \citet{Schafer2025A&A...694A..57S} demonstrate this distinction by measuring the dust dispersion in two ways: using the root-mean-square height and the standard deviation, as shown in the left panel of their Fig.~4. The vertical Reynolds stress, together with the standard deviation $\sigma_{z_\mathrm{d}}$, is more sensitive to small-scale diffusive mixing, whereas the dust scale height, and similarly the root-mean-square height $z_{\mathrm{d,rms}}$, also captures large-scale coherent structures. The VSI, and likewise the DVSI, can therefore produce substantially stronger large-scale vertical transport than small-scale turbulent diffusion. This explains why the diffusion inferred from the tracer scale height is larger than that estimated from the vertical Reynolds stress tensor.


\section{Parameter study}\label{sec:parameterstudy}

\par The DVSI depends only on the vertical shear profile $q_z(z)$, which is determined by several disc parameters, namely the metallicity $Z$, the dust diffusion coefficient $\alpha_D$, the Stokes number $\tau_\mathrm{s}$, and the background pressure gradient $\eta$.


\begin{figure}[ht!]
    \includegraphics[width=\hsize]{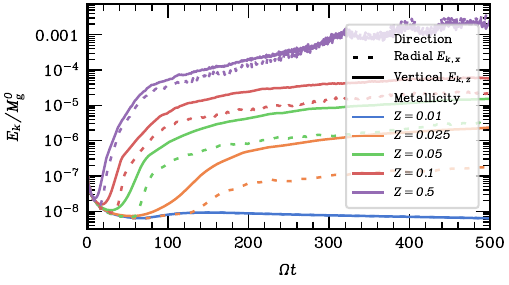}
    \caption{Time evolution of the directional kinetic energy normalised by the initial gas mass in the simulation box for runs \texttt{DVSI}$_{Z,\, .01}$,  \texttt{DVSI}$_{Z,\, .025}$,  \texttt{DVSI}, \texttt{DVSI}$_{Z,\, .1}$ and \texttt{DVSI}$_{Z,\, .5}$.}
    \label{fig:DVSI_Z_Ek}
\end{figure}

\subsection{Metallicity}\label{sec:metallicity}

\par The metallicity and the background pressure gradient primarily affect the amplitude of the vertical shear profile, while the width remains largely unchanged. In this study, we keep the background pressure gradient fixed at $\eta = 0.05$. Increasing the metallicity increases the maximum value of $q_z$, which causes the DVSI to grow faster and reach higher velocities before it is stabilised by the secondary instability. Figure \ref{fig:DVSI_Z_Ek} shows that, from run \texttt{DVSI}$_{Z,\,0.025}$ to run \texttt{DVSI}$_{Z,\,0.5}$, the kinetic energy grows faster and reaches higher saturation levels. At larger metallicities, the difference between the radial and vertical component of the kinetic energy indicates that the turbulence becomes more isotropic, which could be explained by radial KH roll-up, see e.g. \citetalias{Sengupta2023ApJ...942...74S}. Run \texttt{DVSI}$_{Z,\,0.01}$, however, does not grow. In this case, the vertical shear that drives the perturbation is weaker than the numerical viscosity of the simulation, preventing the instability from developing.


\begin{figure}[ht!]
    \includegraphics[width=\hsize]{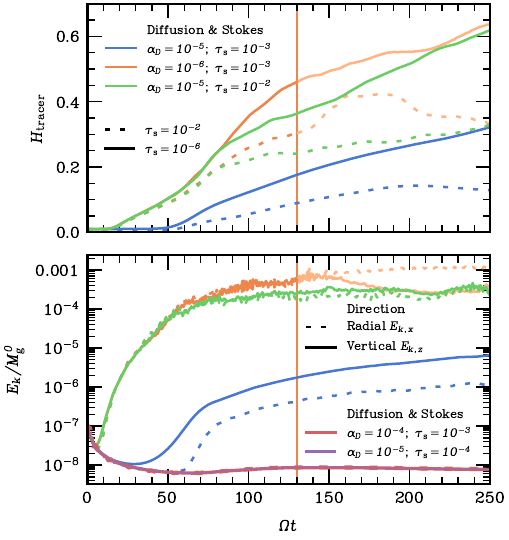}
    \caption{Same as Fig. \ref{fig:DVSI_Ngrid_Htracer_Ek} but for runs \texttt{DVSI}, \texttt{DVSI}$_{\alpha_D,\, 1\mathrm{e}-6}$, \texttt{DVSI}$_{\tau_\mathrm{s}, 1\mathrm{e}-2}$, \texttt{DVSI}$_{\alpha_D,\, 1\mathrm{e}-4}$ and \texttt{DVSI}$_{\tau_\mathrm{s}, 1\mathrm{e}-4}$. The vertical lines show $\Omega t$ after which numerical artefacts appear.}
    \label{fig:DVSI_aD_ts_Htracer_Ek}
\end{figure}

\begin{figure}[ht!]
    \includegraphics[width=\hsize]{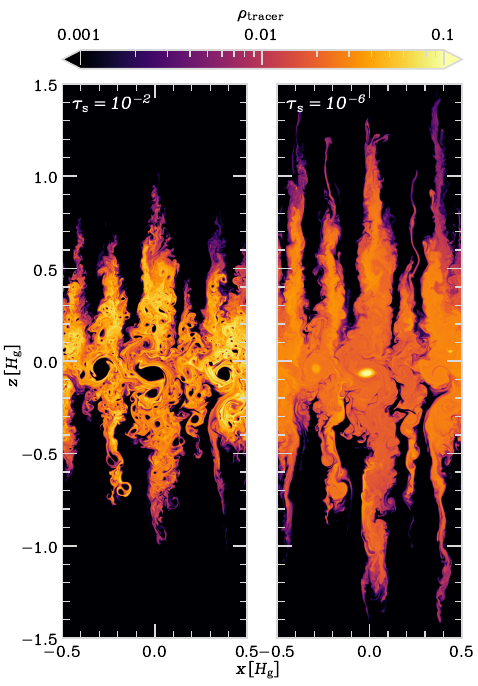}
    \caption{Same as Fig. \ref{fig:DVSI_heatmap_tracer_t500} but for run \texttt{DVSI}$_{\alpha_D,\, 1\mathrm{e}-6}$ at $\Omega t = 125$.}
    \label{fig:DVSI_aD_tracer}
\end{figure}

\subsection{Dust diffusion and Stokes number}\label{sec:dust_diffusion}

\par In the “dust-analogue” approach without dynamical dust evolution, variations in the Stokes number and dust diffusion have very similar effects on the equilibrium dust distribution, as both primarily determine the dust scale height. The scale height is crucial for the DVSI because it sets the gradient of the vertical shear profile: a thinner dust layer produces a narrower shear band with a larger maximum shear. Consequently, increasing the Stokes number or decreasing the dust diffusion strengthens the DVSI.

\par This behaviour is visible in the kinetic energy evolution shown in Fig. \ref{fig:DVSI_aD_ts_Htracer_Ek}. Runs \texttt{DVSI}$_{\alpha_D,\,1\mathrm{e}-6}$ and \texttt{DVSI}$_{\tau_\mathrm{s},\,1\mathrm{e}-2}$ evolve very similarly, both in the tracer scale height (top panel) and in the directional kinetic energy (bottom panel). However, these runs differ from the standard run \texttt{DVSI}. In the standard case, a characteristic feature of the DVSI is that the vertical kinetic energy saturates at a higher level than the radial component. In contrast, in runs \texttt{DVSI}$_{\alpha_D,\,1\mathrm{e}-6}$ and \texttt{DVSI}$_{\tau_\mathrm{s},\,1\mathrm{e}-2}$ the radial and vertical kinetic energy components converge and evolve together (similar to run \texttt{DVSI}$_{Z,\,0.5}$).

\par This behaviour can be explained by the presence of strong isotropic vortices, which dominate over the extended modes typically associated with the VSI and DVSI. In this regime, the dust layer that drives the DVSI is significantly thinner, making the dynamics resemble a KH-like instability, discussed further in Sect.~\ref{sec:KHregime}. The vortices, therefore, become the primary driver of vertical mixing, rather than the thin vertical modes. This is visible in Fig.~\ref{fig:DVSI_aD_tracer}, which shows the density distribution of the two tracers in run \texttt{DVSI}$_{\alpha_D,\,1\mathrm{e}-6}$ at $\Omega t = 175$. Compared to the standard DVSI case (Fig. \ref{fig:DVSI_heatmap_tracer_t500}), no spikes associated with extended modes are present. Instead, mixing is dominated by vortices, and the radial width of the vertical structures is set by the size of the largest vortex. The vortices can expel or concentrate dust, depending on the Stokes Number \citep[see e.g.][]{Cuzzi2001ApJ...546..496C, Johansen2007Natur.448.1022J, Gerosa2023EPJP..138....9G}. In Fig.~\ref{fig:DVSI_aD_tracer}, we can observe that the tracer fluid with the larger Stokes numbers is dispersed by the eddies, while the smaller tracer fluid ($\tau_\mathrm{s}$) is concentrated in the centre of the eddies. The eddies can concentrate the smaller tracers on small scales, but it still diffuse the smaller tracers on larger scales.

\par Runs \texttt{DVSI}$_{\alpha_D,\,1\mathrm{e}-4}$ and \texttt{DVSI}$_{\tau_\mathrm{s},\,1\mathrm{e}-4}$ do not show sustained growth. Similar to run \texttt{DVSI}$_{Z,\,0.01}$, we observe a striped pattern at the grid scale with an amplitude of order $10^{-4}c\mathrm{s}$. In contrast, in runs \texttt{DVSI}$_{\alpha_D,\,1\mathrm{e}-6}$ the turbulence becomes so strong that it appears to trigger thin numerical stripes (the time after which they appear is indicated with a vertical line in Fig. \ref{fig:DVSI_aD_ts_Htracer_Ek}), which are subsequently amplified and produce reflections at the vertical boundaries. The thin numerical stripes and the reflections in the vertical velocity are visible in the appendix Fig.~\ref{fig:DVSI_gas_t500_ad6}. These numerical artefacts in the gas are similar to those reported in Sect.~3.5.5 of \citet{Benitez-Llambay2019ApJS..241...25B}, footnote 6 of \citet{Matthijsse2025A&A...695A.158M}, and Appendix A of \citet{Aly2025A&A...701A.105A}. Although the orientation and the spacing between these thin stripes are more chaotic than that visible in an unstratified shearing box. This does mean that the numerical artefacts can appear in a gas-only simulation run, and in this setup, can not be related to the dust-gas feedback.


\begin{figure}[ht!]
    \includegraphics[width=\hsize]{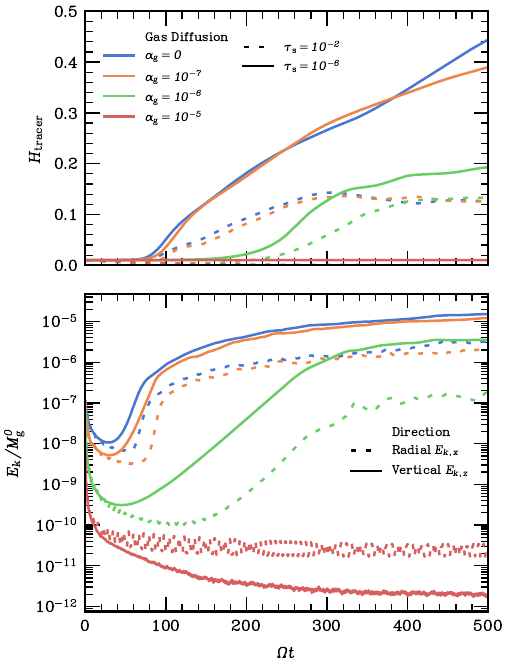}
    \caption{Same as Fig. \ref{fig:DVSI_Ngrid_Htracer_Ek} but for runs \texttt{DVSI}, \texttt{DVSI}$_{\alpha_\mathrm{g},\, 1\mathrm{e}-7}$, \texttt{DVSI}$_{\alpha_\mathrm{g},\, 1\mathrm{e}-6}$, and \texttt{DVSI}$_{\alpha_\mathrm{g},\, 1\mathrm{e}-5}$.}
    \label{fig:DVSI_ag_Htracer_Ek}
\end{figure}

\begin{figure}[ht!]
    \includegraphics[width=\hsize]{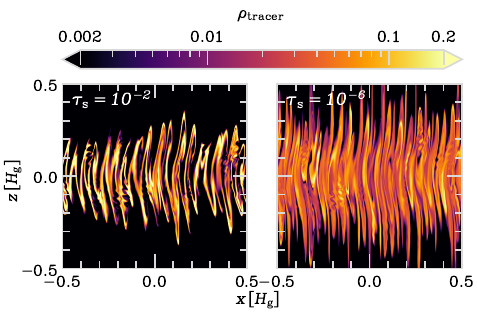}
    \caption{Same as Fig. \ref{fig:DVSI_heatmap_tracer_t500} but for run \texttt{DVSI}$_{\alpha_\mathrm{g},\, 1\mathrm{e}-6}$ at $\Omega t = 500$, and only showing $L_z/8$.}
    \label{fig:DVSI_ag_tracer}
\end{figure}

\subsection{Gas diffusion}\label{sec:gas_diffusion}

\par The inclusion of gas diffusion has an effect on the DVSI similar to that of the grid resolution. At very low gas diffusion coefficients ($\alpha_\mathrm{g} \lesssim 10^{-7}$), the instability is only weakly affected. In Fig.~\ref{fig:DVSI_ag_Htracer_Ek}, run \texttt{DVSI}$_{\alpha_\mathrm{g},\,1\mathrm{e}-7}$ evolves very similarly to the standard run \texttt{DVSI} without gas diffusion. However, if the gas diffusion is too strong ($\alpha_\mathrm{g} \gtrsim 10^{-5}$), the DVSI does not develop.

\par For the intermediate case $\alpha_\mathrm{g} \sim 10^{-6}$, the DVSI initially develops similarly to the case without gas diffusion, albeit with a slightly slower growth rate, and still produces the same extended vertical modes. However, unlike the standard run \texttt{DVSI}, it does not develop the secondary KH-like instability and no eddies form.

\par This significantly affects the vertical mixing. Without the secondary instability, the diffusion of the extended modes is weak, and the initial width of the tracer layer changes only marginally. The tracer fluids leave the midplane and are transported upwards by the vertical modes, but they are not efficiently mixed \citep[similar to the 2D VSI-only cases of][]{Huang2025ApJ...986...76H, Schafer2025A&A...694A..57S}.

\par Comparing the tracer scale heights in Fig.~\ref{fig:DVSI_ag_Htracer_Ek} and the tracer density distributions of run \texttt{DVSI}$_{\alpha_\mathrm{g},\,1\mathrm{e}-6}$ (Fig.~\ref{fig:DVSI_ag_tracer} ) with those of the standard run \texttt{DVSI} (Fig.~\ref{fig:DVSI_heatmap_tracer_t500}), we find that the inclusion of gas diffusion reduces the vertical extent of the strongly coupled tracer ($\tau_\mathrm{s}=10^{-6}$), while the more weakly coupled tracer ($\tau_\mathrm{s}=10^{-2}$) is largely unaffected. This difference can be attributed to the reduced vertical turbulent kinetic energy in run \texttt{DVSI}$_{\alpha_\mathrm{g},\,1\mathrm{e}-6}$, as shown by $E_{k,z}$ in Fig.~\ref{fig:DVSI_ag_Htracer_Ek}. However, if vertical turbulent motions alone determined the tracer scale height, we would also expect a reduction for the $\tau_\mathrm{s}=10^{-2}$ tracer. Instead, its vertical extent remains largely unchanged, although its radial diffusion is reduced. This suggests that the eddies play an important role in redistributing tracer material between vertical channels at higher elevations. In their absence, the tracer cannot readily diffuse radially into neighbouring channels when the vertical velocity of a spike reverses (see Sect.~\ref{sec:vertical_mixning}).


\section{Discussion and conclusion}\label{sec:discussion_conclusion}

\par In this study, we investigated a vertically stratified, isothermal, barotropic shearing box in which the vertical shear is not produced by a global radial temperature gradient, as in the classical VSI, but through the addition of a constant forcing term, that corresponds to a settled, static dust layer that does not evolve. In this setup, the dust layer modifies the azimuthal velocity profile of the gas and provides the free energy for a dust-driven vertical shear instability (DVSI).

\par The DVSI drives robust, vertically extended modes, in the absence of dynamical dust feedback. By using a dust-analogue forcing, we retain the azimuthal component of the vertical shear associated with a stratified dust layer, while removing additional dust-gas effects such as settling, radial drift, streaming-type instabilities, and dusty buoyancy. This makes the DVSI a useful limiting case, as it isolates the gas-dynamical response to the vertical shear produced by the stratified dust layer.

\par In this simplified version of the “dust-analogue” approach, we are able to isolate the DVSI from the SI. However, in stratified SI simulations that include dust feedback, the effects of the DVSI cannot be separated from those of the SI. Moreover, the parameter regimes typically considered in stratified SI simulations are also unstable to the DVSI \citep[see, e.g.,][]{Li2021ApJ...919..107L, Lim2024ApJ...969..130L}.

\par In the linear regime of the DVSI, the perturbations grow fastest in the regions where the vertical shear, $q_z$, is strongest, corresponding to the height at which the dust-to-gas ratio changes most rapidly. The unstable modes are radially short and vertically extended, with large $k_x/k_z$, and grow towards larger heights. This differs from the behaviour often associated with the classical VSI, where surface modes can grow from higher elevations and propagate inwards towards the midplane \citep[see, e.g.][]{Nelson2013MNRAS.435.2610N}. Because the radial wavelength is small, the fastest-growing modes are sensitive to both numerical resolution and gas diffusion. Lower radial resolution, or stronger gas diffusion, damps the large-$k_x$ modes and therefore reduces the measured growth rate.

\par After the linear phase, the vertically extended DVSI modes saturate through a secondary instability. The saturated state is no longer dominated by the coherent, anisotropic modes of the linear phase, but by smaller-scale vortical structures. The morphology of this secondary instability is KH-like: the extended vertical modes generate sufficiently strong radial shear in their vertical velocities to form small, nearly isotropic eddies in the radial-vertical plane. In this respect, the secondary instability of the DVSI is very similar to that of the classical VSI; see Sect. 3.1 of \citet{Latter2018MNRAS.474.3110L}. That the extended modes appear only at larger scale heights makes it important to consider a vertical domain of a stratified simulation that is significantly larger than the dust scale height ($L_z \gg H_\mathrm{d}$).

\subsection{Kelvin-Helmholtz regime}\label{sec:KHregime}

\par It is important to distinguish the parasitic KH-like instability from the classical KHI of dust layers. The simulations in this work are axisymmetric in the radial-vertical plane, so non-axisymmetric perturbations with $k_y \neq 0$ cannot grow. The KH-like behaviour discussed here therefore refers to the secondary saturation mechanism in the radial-vertical plane, not to the classical non-axisymmetric KHI studied in azimuthal-vertical dust-layer configurations \citep[e.g.][]{Johansen2006ApJ...643.1219J, Johansen2006ApJ...636.1121J}.

\par For sufficiently narrow or strong shear layers, the simulations enter a different regime. In this case, the instability is no longer dominated by vertically extended anisotropic modes, but by more isotropic vortices in the radial-vertical plane. This behaviour occurs with the runs that have the largest peak in radial component of the vertical shear profile $\partial_zv_{x}$ (runs \texttt{DVSI}$_{Z,\, 0.5}$, \texttt{DVSI}$_{\alpha_D,\, 1\mathrm{e}-6}$ and \texttt{DVSI}$_{\tau_\mathrm{s}, 1\mathrm{e}-2}$). A possible explanation for this regime change is radial KH roll-up discussed in \citet{Ishitsu2009arXiv0905.4404I} (the $\tau_f\Omega_K=1$ run) and \citetalias{Sengupta2023ApJ...942...74S} (the $\tau_\mathrm{s}=0.2$ runs), where radial drift speed becomes high enough that the system experiences axisymmetric KHI from the radial component of the vertical shear. This provides a practical distinction between the standard DVSI regime, which is characterised by vertically extended modes, and a more KH-like regime, which is dominated by small-scale vortices.

\subsection{Comparison with Lin (2021)}\label{sec:lin2021}

\par In the VSSI, the instability is affected by several coupled dust-gas processes, including dust settling, dust-gas drift, pressure forces, and dusty buoyancy. In contrast, the DVSI simulations from this work remove the dynamic dust component while retaining the vertical shear that would be produced by a stratified dust layer. This means that streaming-related instabilities cannot develop in our setup, because there is no relative dust-gas drift and no drag-mediated feedback loop. In \citetalias{Lin2021ApJ...907...64L}, the dusty buoyancy also produces a strong restoring force that needs to be overcome by the forces driving the instability, such as the vertical shear. Using the “dust analogue” method in an isothermal box, in which dust drag is modelled as a fixed external force, the restoring forces of the dust and gas are completely absent. The DVSI could therefore be interpreted as the purely hydrodynamic vertical-shear component of the VSSI without any contribution from dust settling, dust-gas drift, and the important stabilising effects of dust buoyancy \citepalias{Lin2021ApJ...907...64L}.

\par The relative importance of the instabilities driving mechanisms depends on the radial wavenumber. At low radial wavenumbers, the unstable modes mostly dominated by by a mixutere of relative dust-gas drift. At high radial wavenumbers, the contribution from vertical shear becomes dominant, while dust-gas drift becomes stabilising \citepalias{Lin2021ApJ...907...64L}. The linear analysis of \citetalias{Lin2021ApJ...907...64L} further shows that the growth rates increase with Stokes number and with metallicity (see Figs.~7 and 9 of \citetalias{Lin2021ApJ...907...64L}). Consistent with these trends, our numerical simulations reach saturation more rapidly when either the Stokes number or the metallicity is increased. Finally, we find that gas viscosity begins to stabilise the system for $\alpha_\mathrm{g} \gtrsim 10^{-7}$, in agreement with \citetalias{Lin2021ApJ...907...64L}, who showed that even a viscosity of $\alpha_\mathrm{g}=10^{-7}$ can reduce the linear growth rates.

\subsection{Comparison with Sengupta \& Umurhan (2023)}\label{sec:sengupta2023}

\par The PPD analogue of the baroclinic Symmetric Instability (SymI) proposed in \citetalias{Sengupta2023ApJ...942...74S} works with a single fluid framework and therefore more closely resembles the DVSI and the dust-analogue setup, compared to the VSSI that included the stabilising effect of dust in its two-fluid approach \citepalias{Lin2021ApJ...907...64L}. Both the DVSI and SymI emphasise the same physical process. \citetalias{Sengupta2023ApJ...942...74S} interpret the activity in settled particle layers as arising from the combined action of symmetric instability and KH roll-up, and suggest that the single-fluid theory should be contained within the more general VSSI framework. In this work, we similarly isolate the role of the dust-driven vertical shear by considering a gas-only setup of the VSSI. Within this formulation, the instability is naturally described as a dust-driven vertical shear instability. The close agreement in physical ingredients suggests that the DVSI is not a separate instability from the SymI interpretation, and that both instabilities fall under the more general VSSI framework.

\par There are important differences between the numerical simulation of this work and the 3D axisymmetric simulation of \citetalias{Sengupta2023ApJ...942...74S}. Our runs are gas-only, while the simulation of \citetalias{Sengupta2023ApJ...942...74S} includes super-particles and dust drag. Our simulations are also initialised in an equilibrium state, in a manner similar to \citetalias{Lin2021ApJ...907...64L}, that assumes that there is already a turbulent sublayer where the DVSI develops. This means that there is no initial shock to the system, which allows the linear growth phase of the DVSI to be followed. This is an important distinction from the simulations of \citetalias{Sengupta2023ApJ...942...74S}, whose initial conditions are drawn from actively settling solutions, where there is no dust diffusion. As noted by \citetalias{Sengupta2023ApJ...942...74S}, their results agree in spirit with \citetalias{Lin2021ApJ...907...64L}, but differ in detail because their steady states arise from collapsing dust layers rather than from equilibria constructed using an assumed dust diffusion model. This difference in initialisation may also affect the early-time response of the system. During the initial settling phase, the gas-dust mixture must first adjust towards its quasi-steady state, potentially producing transient shocks or relaxation behaviour before the instability develops (e.g. the Dust Settling Instability \citep[][]{Squire2018MNRAS.477.5011S, Krapp2020MNRAS.497.2715K, Aly2025A&A...701A.105A}). By contrast, our equilibrium setup reduces these initial transients and makes the linear growth of the DVSI visible.

\par A further difference is the size of the vertical domain. The linear modes grow around $1-2 H_\mathrm{d}$ and agree with the analytical analysis of \citetalias{Sengupta2023ApJ...942...74S}. The saturated nonlinear modes extend way higher up to heights of order the gas scale height, $\mathcal{O}(H_\mathrm{g})$, which allows us to capture the vertically extended behaviour of the instability at higher altitudes. In these regions, the perturbations show a VSI-like vertical structure, with modes extending well above the dense midplane dust layer. Such behaviour is not captured in simulations with a smaller vertical domain, such as those of \citet{Ishitsu2009arXiv0905.4404I} and \citetalias{Sengupta2023ApJ...942...74S}. The larger vertical box, therefore, reveals that the DVSI is not confined to the immediate dust layer, but can excite extended gas motions throughout a substantial fraction of the disc. Although some of the VSI-like structures are visible in isothermal runs with dust feedback, see e.g. Fig.~3 from \citet{Lehmann2025ApJ...983..135L}\footnote{This run is not isothermal but approaches the isothermal limit with $\beta=10^{-3}$.}, Fig.~9 from \citet{Tatarelli2026A&A...711A.252T} and \citetalias{MatthijsseInprepA&APaperII}.

\subsection{Further work}\label{sec:further_work}

\par The stabilising effect of dust feedback is investigated in \citetalias{MatthijsseInprepA&APaperII}, where the DVSI with dust feedback (equivalent to the VSSI) grows significantly slower than without dust feedback. The dust feedback acts as a stabilising force on the VSSI, through dusty buoyancy and increased inertia \citep[see,][and \citetalias{Lin2021ApJ...907...64L}]{Lin2017ApJ...849..129L, Lin2019MNRAS.485.5221L}. However, this stabilising effect of the feedback can be weakened by changing the dust layer that drives the DVSI; if this layer has a combination of Stokes number and dust diffusion that is less coupled to the gas, the stabilising effect becomes significantly smaller compared to the vertical shear. We therefore also investigate if the DVSI can grow and can drive vertical transport that corresponds to constraints from observations in dust traps like IRS-48 \citep[][]{vandermarel_A_2021A&A...651L...5V, Leemker2023A&A...673A...7L, Temmink2025A&A...693A.101T}. Because in these regions there are strong constraints on vertical mixing time scale, and dust-rich regions like dust traps are regions where the “classical” VSI could be stabilised by the dust \citep[see e.g.][]{Lin2017ApJ...849..129L, Lin2019MNRAS.485.5221L, Schafer2020A&A...635A.190S, Schafer2025A&A...694A..57S, Fukuhara2021ApJ...914..132F, Fukuhara2023PASJ...75..233F, Fukuhara2024PASJ...76..708F, Fukuhara2025A&A...701A..72F, Lehmann2022A&A...658A.156L, Lehmann2023MNRAS.522.5892L, Pfeil2023ApJ...959..121P, Pfeil2024A&A...687L...5P}.

\par With the local setup, there is the possibility to add a vertically dependent acceleration field that more closely resembles the global temperature structure in a local setup \citep[similar to analytical work of][]{McNally2015ApJ...811..121M}, allowing for much higher resolution simulations than in radially global setups of the classical VSI. The radially unstratified setup with an imposed vertical shear might also be a good setup to investigate the interplay between the VSI and SI \citep[e.g.][]{Huang2025ApJ...986...76H, Schafer2025A&A...694A..57S}.

\par The parameters for the runs lie inside the strong clumping criteria for the stratified turbulent SI determined by \citet{Lim2025ApJ...981..160L}, but these criteria are without dust diffusion, and the presence of external turbulence can significantly change the strong clumping criteria, see \citet{Lim2024ApJ...969..130L}. In further work, the parameter space can be extended to investigate how strong the DVSI plays in the stratified SI runs where there is strong clumping \citep[see also][]{Lim2025ApJ...993...12L}. The strength of dust diffusion in our runs is strong and could prevent strong clumping through the SI \citep[][]{Gole2020ApJ...904..132G,  Lim2024ApJ...969..130L, Matthijsse2025A&A...695A.158M}.

\par We only considered two spatial dimensions; this meant that we did not have any non-axisymmetric effect like the classical KHI. It is important to investigate the non-axisymmetric effect in further work. The secondary instability seen near the midplane could be exaggerated in axisymmetric runs. This is because non-axisymmetric disturbances are sheared out on their instability timescale and only grow transiently \citep[][]{Latter2018MNRAS.474.3110L}, and the strong eddies in the azimuthal plane are not visible in 3D VSI simulations \citep[see e.g.][]{Lesur2025A&A...703A.225L}. The dustlayer driving the DVSI is also monodisperse; if we take into account a polydisperse setup, the different sizes will have different equilibrium dust scale heights, affecting the combined dust shear profile.


\subsection{Summary}\label{sec:summery}

\par We investigated a dust-driven instability in an isothermal shearing box. This instability is driven by vertical shear, where the shear profile comes from a settled dust layer instead of a global barotropic temperature structure.

\begin{itemize} 

    \item In the linear regime, we observe S-shaped modes in the velocity perturbations, which are radially narrow and vertically extended (large $k_x/k_z$). These modes originate at a height where the vertical shear profile peaks. This corresponds to the height at which the dust-to-gas ratio changes most.

    \item These extended modes grow outwards to larger $z$ values and also cross midplane, creating vertically extended anisotropic modes that are stabilised by a secondary instability, creating small-scale vortices that are isotropic. It is therefore important to explore simulations with a large vertical extent and capture the large vertical modes of the DVSI ($L_z \gg H_\mathrm{d}$).

    \item Dust profiles of the type considered in stratified SI studies \citep[see e.g.][]{Li2021ApJ...919..107L, Lim2024ApJ...969..130L} are likely to be susceptible to the DVSI. The dust-induced vertical shear may therefore drive turbulence alongside, or prior to, the development of stratified SI modes.

    \item If the vertical shear profile becomes narrow and has a higher amplitude (very settled dust layer), there is a radial KH roll-up regime, where the turbulence driven by the shear is no longer anisotropic (extended vertical modes) but dominated by isotropic vortices in the radial vertical plane.

    \item Lower resolution or gas diffusion dampens the largest wavenumbers (which are also the fastest growing modes) of the DVSI and secondary parasitic instability, decreasing the growth rate of the DVSI and quenching the radial (small-scale) diffusion of the secondary instability.
\end{itemize}


Data Availability: The data that support the findings of this study are openly available in 4TU.ResearchData under the name: Data underlying the publication: “Dust-driven vertical shear instability in an isothermal shearing box” at \url{http://doi.org/10.4121/571e09ec-10db-452d-b541-b7d86f573e50}. The code that processes the output of \texttt{FARGO3D} and supports the findings of this study is openly available in 4TU.ResearchData under the name: Code underlying the publications “Dust-driven vertical shear instability in an isothermal shearing box” at \url{http://doi.org/10.4121/0ab6e524-0c51-4c40-986e-fd5bca604953}.

\begin{acknowledgements}
This project has received funding from the European Research Council (ERC) under the European Union’s Horizon Europe research and innovation programme (Grant Agreement No. 101054502). RPN is supported by STFC grants ST/X000931/1 and UKRI4119. The authors acknowledge the use of computational resources of the DelftBlue supercomputer, provided by \citet{DHPC2024}. This work made use of several open-source software packages. We acknowledge \texttt{numpy} \citep{Harris2020Natur.585..357H}, \texttt{matplotlib} \citep{Hunter2007CSE.....9...90H}, \texttt{scipy} \citep{Virtanen2020NatMe..17..261V}, \texttt{FARGO3D} \citep{Benitez-Llambay2016ApJS..223...11B, Benitez-Llambay2019ApJS..241...25B}, and \texttt{psitools} \citep{Paardekooper2020MNRAS.499.4223P, Paardekooper2021MNRAS.502.1579P, McNally2021MNRAS.502.1469M}. We acknowledge 4TU.ResearchData for supporting open access to research data.
\end{acknowledgements}

\bibliographystyle{aa} 
\bibliography{aa_DVSI_I.bib} 

@ARTICLE{Aly2025A&A...701A.105A,
       author = {{Aly}, Hossam and {Paardekooper}, Sijme-Jan},
        title = "{Non-linear evolution of the unstratified polydisperse dust settling instability}",
      journal = {\aap},
         year = 2025,
        month = sep,
       volume = {701},
          eid = {A105},
        pages = {A105},
          doi = {10.1051/0004-6361/202554415},
archivePrefix = {arXiv},
       eprint = {2507.10719},
 primaryClass = {astro-ph.EP},
       adsurl = {https://ui.adsabs.harvard.edu/abs/2025A&A...701A.105A}
}

@ARTICLE{Bai2010ApJ...722.1437B,
       author = {{Bai}, Xue-Ning and {Stone}, James M.},
        title = "{Dynamics of Solids in the Midplane of Protoplanetary Disks: Implications for Planetesimal Formation}",
      journal = {\apj},
         year = 2010,
        month = oct,
       volume = {722},
       number = {2},
        pages = {1437-1459},
          doi = {10.1088/0004-637X/722/2/1437},
archivePrefix = {arXiv},
       eprint = {1005.4982},
 primaryClass = {astro-ph.EP},
       adsurl = {https://ui.adsabs.harvard.edu/abs/2010ApJ...722.1437B}
}

@ARTICLE{Balbus1991ApJ...376..214B,
       author = {{Balbus}, Steven A. and {Hawley}, John F.},
        title = "{A Powerful Local Shear Instability in Weakly Magnetized Disks. I. Linear Analysis}",
      journal = {\apj},
         year = 1991,
        month = jul,
       volume = {376},
        pages = {214},
          doi = {10.1086/170270},
       adsurl = {https://ui.adsabs.harvard.edu/abs/1991ApJ...376..214B}
}

@ARTICLE{Barker2015MNRAS.450...21B,
       author = {{Barker}, A.~J. and {Latter}, H.~N.},
        title = "{On the vertical-shear instability in astrophysical discs}",
      journal = {\mnras},
         year = 2015,
        month = jun,
       volume = {450},
       number = {1},
        pages = {21-37},
          doi = {10.1093/mnras/stv640},
archivePrefix = {arXiv},
       eprint = {1503.06953},
 primaryClass = {astro-ph.SR},
       adsurl = {https://ui.adsabs.harvard.edu/abs/2015MNRAS.450...21B}
}

@ARTICLE{Benitez-Llambay2016ApJS..223...11B,
       author = {{Ben{\'\i}tez-Llambay}, Pablo and {Masset}, Fr{\'e}d{\'e}ric S.},
        title = "{FARGO3D: A New GPU-oriented MHD Code}",
      journal = {\apjs},
         year = 2016,
        month = mar,
       volume = {223},
       number = {1},
          eid = {11},
        pages = {11},
          doi = {10.3847/0067-0049/223/1/11},
archivePrefix = {arXiv},
       eprint = {1602.02359},
 primaryClass = {astro-ph.IM},
       adsurl = {https://ui.adsabs.harvard.edu/abs/2016ApJS..223...11B}
}

@ARTICLE{Benitez-Llambay2019ApJS..241...25B,
       author = {{Ben{\'\i}tez-Llambay}, Pablo and {Krapp}, Leonardo and {Pessah}, Martin E.},
        title = "{Asymptotically Stable Numerical Method for Multispecies Momentum Transfer: Gas and Multifluid Dust Test Suite and Implementation in FARGO3D}",
      journal = {\apjs},
         year = 2019,
        month = apr,
       volume = {241},
       number = {2},
          eid = {25},
        pages = {25},
          doi = {10.3847/1538-4365/ab0a0e},
archivePrefix = {arXiv},
       eprint = {1811.07925},
 primaryClass = {astro-ph.EP},
       adsurl = {https://ui.adsabs.harvard.edu/abs/2019ApJS..241...25B}
}

@ARTICLE{Birnstiel2012A&A...539A.148B,
       author = {{Birnstiel}, T. and {Klahr}, H. and {Ercolano}, B.},
        title = "{A simple model for the evolution of the dust population in protoplanetary disks}",
      journal = {\aap},
         year = 2012,
        month = mar,
       volume = {539},
          eid = {A148},
        pages = {A148},
          doi = {10.1051/0004-6361/201118136},
archivePrefix = {arXiv},
       eprint = {1201.5781},
 primaryClass = {astro-ph.EP},
       adsurl = {https://ui.adsabs.harvard.edu/abs/2012A&A...539A.148B}
}

@ARTICLE{Chen2020ApJ...891..132C,
       author = {{Chen}, Kan and {Lin}, Min-Kai},
        title = "{How Efficient Is the Streaming Instability in Viscous Protoplanetary Disks?}",
      journal = {\apj},
         year = 2020,
        month = mar,
       volume = {891},
       number = {2},
          eid = {132},
        pages = {132},
          doi = {10.3847/1538-4357/ab76ca},
archivePrefix = {arXiv},
       eprint = {2002.07188},
 primaryClass = {astro-ph.EP},
       adsurl = {https://ui.adsabs.harvard.edu/abs/2020ApJ...891..132C}
}

@ARTICLE{Cuzzi1993Icar..106..102C,
       author = {{Cuzzi}, Jeffrey N. and {Dobrovolskis}, Anthony R. and {Champney}, Joelle M.},
        title = "{Particle-Gas Dynamics in the Midplane of a Protoplanetary Nebula}",
      journal = {\icarus},
         year = 1993,
        month = nov,
       volume = {106},
       number = {1},
        pages = {102-134},
          doi = {10.1006/icar.1993.1161},
       adsurl = {https://ui.adsabs.harvard.edu/abs/1993Icar..106..102C}
}

@ARTICLE{Cuzzi2001ApJ...546..496C,
       author = {{Cuzzi}, Jeffrey N. and {Hogan}, Robert C. and {Paque}, Julie M. and {Dobrovolskis}, Anthony R.},
        title = "{Size-selective Concentration of Chondrules and Other Small Particles in Protoplanetary Nebula Turbulence}",
      journal = {\apj},
         year = 2001,
        month = jan,
       volume = {546},
       number = {1},
        pages = {496-508},
          doi = {10.1086/318233},
archivePrefix = {arXiv},
       eprint = {astro-ph/0009210},
 primaryClass = {astro-ph},
       adsurl = {https://ui.adsabs.harvard.edu/abs/2001ApJ...546..496C}
}

@ARTICLE{Dubrulle1995Icar..114..237D,
       author = {{Dubrulle}, B. and {Morfill}, G. and {Sterzik}, M.},
        title = "{The dust subdisk in the protoplanetary nebula.}",
      journal = {\icarus},
         year = 1995,
        month = apr,
       volume = {114},
       number = {2},
        pages = {237-246},
          doi = {10.1006/icar.1995.1058},
       adsurl = {https://ui.adsabs.harvard.edu/abs/1995Icar..114..237D}
}

@ARTICLE{Flores2020A&A...644A..50F,
       author = {{Flores-Rivera}, Lizxandra and {Flock}, Mario and {Nakatani}, Riouhei},
        title = "{Hydrodynamical simulations of protoplanetary disks including irradiation of stellar photons. I. Resolution study for vertical shear instability}",
      journal = {\aap},
         year = 2020,
        month = dec,
       volume = {644},
          eid = {A50},
        pages = {A50},
          doi = {10.1051/0004-6361/202039294},
archivePrefix = {arXiv},
       eprint = {2010.06711},
 primaryClass = {astro-ph.EP},
       adsurl = {https://ui.adsabs.harvard.edu/abs/2020A&A...644A..50F}
}

@ARTICLE{Fricke1968ZA.....68..317F,
       author = {{Fricke}, K.},
        title = "{Instabilit{\"a}t station{\"a}rer Rotation in Sternen}",
      journal = {\zap},
         year = 1968,
        month = jan,
       volume = {68},
        pages = {317},
       adsurl = {https://ui.adsabs.harvard.edu/abs/1968ZA.....68..317F}
}

@ARTICLE{Fukuhara2021ApJ...914..132F,
       author = {{Fukuhara}, Yuya and {Okuzumi}, Satoshi and {Ono}, Tomohiro},
        title = "{Effects of Dust Evolution on the Vertical Shear Instability in the Outer Regions of Protoplanetary Disks}",
      journal = {\apj},
         year = 2021,
        month = jun,
       volume = {914},
       number = {2},
          eid = {132},
        pages = {132},
          doi = {10.3847/1538-4357/abfe5c},
archivePrefix = {arXiv},
       eprint = {2105.02403},
 primaryClass = {astro-ph.EP},
       adsurl = {https://ui.adsabs.harvard.edu/abs/2021ApJ...914..132F}
}

@ARTICLE{Fukuhara2023PASJ...75..233F,
       author = {{Fukuhara}, Yuya and {Okuzumi}, Satoshi and {Ono}, Tomohiro},
        title = "{Two saturated states of the vertical shear instability in protoplanetary disks with vertically varying cooling times}",
      journal = {\pasj},
         year = 2023,
        month = feb,
       volume = {75},
       number = {1},
        pages = {233-249},
          doi = {10.1093/pasj/psac107},
archivePrefix = {arXiv},
       eprint = {2212.02743},
 primaryClass = {astro-ph.EP},
       adsurl = {https://ui.adsabs.harvard.edu/abs/2023PASJ...75..233F}
}

@ARTICLE{Fukuhara2024PASJ...76..708F,
       author = {{Fukuhara}, Yuya and {Okuzumi}, Satoshi},
        title = "{A self-consistent model for dust settling and the vertical shear instability in protoplanetary disks}",
      journal = {\pasj},
         year = 2024,
        month = aug,
       volume = {76},
       number = {4},
        pages = {708-719},
          doi = {10.1093/pasj/psae042},
archivePrefix = {arXiv},
       eprint = {2404.15780},
 primaryClass = {astro-ph.EP},
       adsurl = {https://ui.adsabs.harvard.edu/abs/2024PASJ...76..708F}
}

@ARTICLE{Fukuhara2025A&A...701A..72F,
       author = {{Fukuhara}, Y. and {Flock}, M. and {Okuzumi}, S. and {Tominaga}, R.~T.},
        title = "{Hydrodynamical simulations of the vertical shear instability with dynamic dust and cooling rates in protoplanetary disks}",
      journal = {\aap},
         year = 2025,
        month = sep,
       volume = {701},
          eid = {A72},
        pages = {A72},
          doi = {10.1051/0004-6361/202555624},
archivePrefix = {arXiv},
       eprint = {2507.11156},
 primaryClass = {astro-ph.EP},
       adsurl = {https://ui.adsabs.harvard.edu/abs/2025A&A...701A..72F}
}

@ARTICLE{Gammie1996ApJ...457..355G,
       author = {{Gammie}, Charles F.},
        title = "{Layered Accretion in T Tauri Disks}",
      journal = {\apj},
         year = 1996,
        month = jan,
       volume = {457},
        pages = {355},
          doi = {10.1086/176735},
       adsurl = {https://ui.adsabs.harvard.edu/abs/1996ApJ...457..355G}
}

@ARTICLE{Gerosa2023EPJP..138....9G,
       author = {{Gerosa}, Fabiola A. and {M{\'e}heut}, H{\'e}lo{\"\i}se and {Bec}, J{\'e}r{\'e}mie},
        title = "{Clusters of heavy particles in two-dimensional Keplerian turbulence}",
      journal = {European Physical Journal Plus},
         year = 2023,
        month = jan,
       volume = {138},
       number = {1},
          eid = {9},
        pages = {9},
          doi = {10.1140/epjp/s13360-022-03585-8},
archivePrefix = {arXiv},
       eprint = {2210.13147},
 primaryClass = {astro-ph.EP},
       adsurl = {https://ui.adsabs.harvard.edu/abs/2023EPJP..138....9G}
}

@ARTICLE{Goldreich1965MNRAS.130..125G,
       author = {{Goldreich}, P. and {Lynden-Bell}, D.},
        title = "{II. Spiral arms as sheared gravitational instabilities}",
      journal = {\mnras},
         year = 1965,
        month = jan,
       volume = {130},
        pages = {125},
          doi = {10.1093/mnras/130.2.125},
       adsurl = {https://ui.adsabs.harvard.edu/abs/1965MNRAS.130..125G}
}

@ARTICLE{Goldreich1967ApJ...150..571G,
       author = {{Goldreich}, Peter and {Schubert}, Gerald},
        title = "{Differential Rotation in Stars}",
      journal = {\apj},
         year = 1967,
        month = nov,
       volume = {150},
        pages = {571},
          doi = {10.1086/149360},
       adsurl = {https://ui.adsabs.harvard.edu/abs/1967ApJ...150..571G}
}

@ARTICLE{Gole2020ApJ...904..132G,
       author = {{Gole}, Daniel A. and {Simon}, Jacob B. and {Li}, Rixin and {Youdin}, Andrew N. and {Armitage}, Philip J.},
        title = "{Turbulence Regulates the Rate of Planetesimal Formation via Gravitational Collapse}",
      journal = {\apj},
         year = 2020,
        month = dec,
       volume = {904},
       number = {2},
          eid = {132},
        pages = {132},
          doi = {10.3847/1538-4357/abc334},
archivePrefix = {arXiv},
       eprint = {2001.10000},
 primaryClass = {astro-ph.EP},
       adsurl = {https://ui.adsabs.harvard.edu/abs/2020ApJ...904..132G}
}

@ARTICLE{Harris2020Natur.585..357H,
       author = {{Harris}, Charles R. and {Millman}, K. Jarrod and {van der Walt}, St{\'e}fan J. and {Gommers}, Ralf and {Virtanen}, Pauli and {Cournapeau}, David and {Wieser}, Eric and {Taylor}, Julian and {Berg}, Sebastian and {Smith}, Nathaniel J. and {Kern}, Robert and {Picus}, Matti and {Hoyer}, Stephan and {van Kerkwijk}, Marten H. and {Brett}, Matthew and {Haldane}, Allan and {del R{\'\i}o}, Jaime Fern{\'a}ndez and {Wiebe}, Mark and {Peterson}, Pearu and {G{\'e}rard-Marchant}, Pierre and {Sheppard}, Kevin and {Reddy}, Tyler and {Weckesser}, Warren and {Abbasi}, Hameer and {Gohlke}, Christoph and {Oliphant}, Travis E.},
        title = "{Array programming with NumPy}",
      journal = {\nat},
         year = 2020,
        month = sep,
       volume = {585},
       number = {7825},
        pages = {357-362},
          doi = {10.1038/s41586-020-2649-2},
archivePrefix = {arXiv},
       eprint = {2006.10256},
 primaryClass = {cs.MS},
       adsurl = {https://ui.adsabs.harvard.edu/abs/2020Natur.585..357H}
}

@ARTICLE{Heinzeller2011ApJ...731..115H,
       author = {{Heinzeller}, D. and {Nomura}, H. and {Walsh}, C. and {Millar}, T.~J.},
        title = "{Chemical Evolution of Protoplanetary Disks{\textemdash}The Effects of Viscous Accretion, Turbulent Mixing, and Disk Winds}",
      journal = {\apj},
         year = 2011,
        month = apr,
       volume = {731},
       number = {2},
          eid = {115},
        pages = {115},
          doi = {10.1088/0004-637X/731/2/115},
archivePrefix = {arXiv},
       eprint = {1102.3972},
 primaryClass = {astro-ph.EP},
       adsurl = {https://ui.adsabs.harvard.edu/abs/2011ApJ...731..115H}
}

@ARTICLE{Huang2025ApJ...986...76H,
       author = {{Huang}, Pinghui and {Bai}, Xue-Ning},
        title = "{The Interplay between Dust Dynamics and Turbulence Induced by the Vertical Shear Instability}",
      journal = {\apj},
         year = 2025,
        month = jun,
       volume = {986},
       number = {1},
          eid = {76},
        pages = {76},
          doi = {10.3847/1538-4357/add345},
archivePrefix = {arXiv},
       eprint = {2503.01656},
 primaryClass = {astro-ph.EP},
       adsurl = {https://ui.adsabs.harvard.edu/abs/2025ApJ...986...76H}
}

@ARTICLE{Hunter2007CSE.....9...90H,
       author = {{Hunter}, John D.},
        title = "{Matplotlib: A 2D Graphics Environment}",
      journal = {Computing in Science and Engineering},
         year = 2007,
        month = may,
       volume = {9},
       number = {3},
        pages = {90-95},
          doi = {10.1109/MCSE.2007.55},
       adsurl = {https://ui.adsabs.harvard.edu/abs/2007CSE.....9...90H}
}

@ARTICLE{Ishitsu2009arXiv0905.4404I,
       author = {{Ishitsu}, Naoki and {Inutsuka}, Shu-ichiro and {Sekiya}, Minoru},
        title = "{Two-fluid Instability of Dust and Gas in the Dust Layer of a Protoplanetary Disk}",
      journal = {arXiv e-prints},
         year = 2009,
        month = may,
          eid = {arXiv:0905.4404},
        pages = {arXiv:0905.4404},
          doi = {10.48550/arXiv.0905.4404},
archivePrefix = {arXiv},
       eprint = {0905.4404},
 primaryClass = {astro-ph.EP},
       adsurl = {https://ui.adsabs.harvard.edu/abs/2009arXiv0905.4404I}
}

@ARTICLE{Johansen2006ApJ...636.1121J,
       author = {{Johansen}, A. and {Klahr}, H. and {Henning}, Th.},
        title = "{Gravoturbulent Formation of Planetesimals}",
      journal = {\apj},
         year = 2006,
        month = jan,
       volume = {636},
       number = {2},
        pages = {1121-1134},
          doi = {10.1086/498078},
       adsurl = {https://ui.adsabs.harvard.edu/abs/2006ApJ...636.1121J}
}

@ARTICLE{Johansen2006ApJ...643.1219J,
       author = {{Johansen}, Anders and {Henning}, Thomas and {Klahr}, Hubert},
        title = "{Dust Sedimentation and Self-sustained Kelvin-Helmholtz Turbulence in Protoplanetary Disk Midplanes}",
      journal = {\apj},
         year = 2006,
        month = jun,
       volume = {643},
       number = {2},
        pages = {1219-1232},
          doi = {10.1086/502968},
archivePrefix = {arXiv},
       eprint = {astro-ph/0512272},
 primaryClass = {astro-ph},
       adsurl = {https://ui.adsabs.harvard.edu/abs/2006ApJ...643.1219J}
}

@ARTICLE{Johansen2007ApJ...662..627J,
       author = {{Johansen}, A. and {Youdin}, A.},
        title = "{Protoplanetary Disk Turbulence Driven by the Streaming Instability: Nonlinear Saturation and Particle Concentration}",
      journal = {\apj},
         year = 2007,
        month = jun,
       volume = {662},
       number = {1},
        pages = {627-641},
          doi = {10.1086/516730},
archivePrefix = {arXiv},
       eprint = {astro-ph/0702626},
 primaryClass = {astro-ph},
       adsurl = {https://ui.adsabs.harvard.edu/abs/2007ApJ...662..627J}
}

@ARTICLE{Johansen2007Natur.448.1022J,
       author = {{Johansen}, Anders and {Oishi}, Jeffrey S. and {Mac Low}, Mordecai-Mark and {Klahr}, Hubert and {Henning}, Thomas and {Youdin}, Andrew},
        title = "{Rapid planetesimal formation in turbulent circumstellar disks}",
      journal = {\nat},
         year = 2007,
        month = aug,
       volume = {448},
       number = {7157},
        pages = {1022-1025},
          doi = {10.1038/nature06086},
archivePrefix = {arXiv},
       eprint = {0708.3890},
 primaryClass = {astro-ph},
       adsurl = {https://ui.adsabs.harvard.edu/abs/2007Natur.448.1022J}
}

@ARTICLE{Krapp2020MNRAS.497.2715K,
       author = {{Krapp}, Leonardo and {Youdin}, Andrew N. and {Kratter}, Kaitlin M. and {Ben{\'\i}tez-Llambay}, Pablo},
        title = "{Dust settling instability in protoplanetary discs}",
      journal = {\mnras},
         year = 2020,
        month = sep,
       volume = {497},
       number = {3},
        pages = {2715-2729},
          doi = {10.1093/mnras/staa1854},
archivePrefix = {arXiv},
       eprint = {2004.04590},
 primaryClass = {astro-ph.EP},
       adsurl = {https://ui.adsabs.harvard.edu/abs/2020MNRAS.497.2715K}
}

@ARTICLE{Latter2006MNRAS.372.1829L,
       author = {{Latter}, Henrik N. and {Ogilvie}, Gordon I.},
        title = "{Viscous overstability and eccentricity evolution in three-dimensional gaseous discs}",
      journal = {\mnras},
         year = 2006,
        month = nov,
       volume = {372},
       number = {4},
        pages = {1829-1839},
          doi = {10.1111/j.1365-2966.2006.11014.x},
archivePrefix = {arXiv},
       eprint = {astro-ph/0609158},
 primaryClass = {astro-ph},
       adsurl = {https://ui.adsabs.harvard.edu/abs/2006MNRAS.372.1829L}
}

@ARTICLE{Latter2018MNRAS.474.3110L,
       author = {{Latter}, Henrik N. and {Papaloizou}, John},
        title = "{Vortices and the saturation of the vertical shear instability in protoplanetary discs}",
      journal = {\mnras},
         year = 2018,
        month = mar,
       volume = {474},
       number = {3},
        pages = {3110-3124},
          doi = {10.1093/mnras/stx3031},
archivePrefix = {arXiv},
       eprint = {1711.07436},
 primaryClass = {astro-ph.EP},
       adsurl = {https://ui.adsabs.harvard.edu/abs/2018MNRAS.474.3110L}
}

@ARTICLE{Leemker2023A&A...673A...7L,
       author = {{Leemker}, M. and {Booth}, A.~S. and {van Dishoeck}, E.~F. and {van der Marel}, N. and {Tabone}, B. and {Ligterink}, N.~F.~W. and {Brunken}, N.~G.~C. and {Hogerheijde}, M.~R.},
        title = "{A major asymmetric ice trap in a planet-forming disk. IV. Nitric oxide gas and a lack of CN tracing sublimating ices and a C/O ratio <1}",
      journal = {\aap},
         year = 2023,
        month = may,
       volume = {673},
          eid = {A7},
        pages = {A7},
          doi = {10.1051/0004-6361/202245662},
archivePrefix = {arXiv},
       eprint = {2303.00768},
 primaryClass = {astro-ph.EP},
       adsurl = {https://ui.adsabs.harvard.edu/abs/2023A&A...673A...7L}
}

@ARTICLE{Lehmann2022A&A...658A.156L,
       author = {{Lehmann}, M. and {Lin}, M. -K.},
        title = "{Impact of local pressure enhancements on dust concentration in turbulent protoplanetary discs}",
      journal = {\aap},
         year = 2022,
        month = feb,
       volume = {658},
          eid = {A156},
        pages = {A156},
          doi = {10.1051/0004-6361/202142378},
archivePrefix = {arXiv},
       eprint = {2112.06153},
 primaryClass = {astro-ph.EP},
       adsurl = {https://ui.adsabs.harvard.edu/abs/2022A&A...658A.156L}
}

@ARTICLE{Lehmann2023MNRAS.522.5892L,
       author = {{Lehmann}, Marius and {Lin}, Min-Kai},
        title = "{Instabilities in dusty non-isothermal protoplanetary discs}",
      journal = {\mnras},
         year = 2023,
        month = jul,
       volume = {522},
       number = {4},
        pages = {5892-5930},
          doi = {10.1093/mnras/stad1349},
archivePrefix = {arXiv},
       eprint = {2305.02362},
 primaryClass = {astro-ph.EP},
       adsurl = {https://ui.adsabs.harvard.edu/abs/2023MNRAS.522.5892L}
}

@ARTICLE{Lehmann2025ApJ...983..135L,
       author = {{Lehmann}, Marius and {Lin}, Min-Kai},
        title = "{Convective Overstability in Radially Global Protoplanetary Disks. II. Impact on Planetesimal Formation}",
      journal = {\apj},
         year = 2025,
        month = apr,
       volume = {983},
       number = {2},
          eid = {135},
        pages = {135},
          doi = {10.3847/1538-4357/adbf06},
archivePrefix = {arXiv},
       eprint = {2503.02084},
 primaryClass = {astro-ph.EP},
       adsurl = {https://ui.adsabs.harvard.edu/abs/2025ApJ...983..135L}
}

@ARTICLE{Lesur2023ASPC..534..465L,
       author = {{Lesur}, G. and {Flock}, M. and {Ercolano}, B. and {Lin}, M.-K. and {Yang}, C. and {Barranco}, J.~A. and {Benitez-Llambay}, P. and {Goodman}, J. and {Johansen}, A. and {Klahr}, H. and {Laibe}, G. and {Lyra}, W. and {Marcus}, P.~S. and {Nelson}, R.~P. and {Squire}, J. and {Simon}, J.~B. and {Turner}, N.~J. and {Umurhan}, O.~M. and {Youdin}, A.~N.},
        title = "{Hydro-, Magnetohydro-, and Dust-Gas Dynamics of Protoplanetary Disks}",
      journal = {arXiv e-prints},
    booktitle = {Protostars and Planets VII},
         year = 2023,
       editor = {{Inutsuka}, S. and {Aikawa}, Y. and {Muto}, T. and {Tomida}, K. and {Tamura}, M.},
       series = {Astronomical Society of the Pacific Conference Series},
       volume = {534},
        month = jul,
        pages = {465},
          doi = {10.48550/arXiv.2203.09821},
archivePrefix = {arXiv},
       eprint = {2203.09821},
 primaryClass = {astro-ph.EP},
       adsurl = {https://ui.adsabs.harvard.edu/abs/2023ASPC..534..465L}
}

@ARTICLE{Lesur2025A&A...703A.225L,
       author = {{Lesur}, G. and {Latter}, H. and {Ogilvie}, G.~I.},
        title = "{High-resolution models of the vertical shear instability}",
      journal = {\aap},
         year = 2025,
        month = nov,
       volume = {703},
          eid = {A225},
        pages = {A225},
          doi = {10.1051/0004-6361/202555944},
archivePrefix = {arXiv},
       eprint = {2508.20839},
 primaryClass = {astro-ph.EP},
       adsurl = {https://ui.adsabs.harvard.edu/abs/2025A&A...703A.225L}
}

@ARTICLE{Li2021ApJ...919..107L,
       author = {{Li}, Rixin and {Youdin}, Andrew N.},
        title = "{Thresholds for Particle Clumping by the Streaming Instability}",
      journal = {\apj},
         year = 2021,
        month = oct,
       volume = {919},
       number = {2},
          eid = {107},
        pages = {107},
          doi = {10.3847/1538-4357/ac0e9f},
archivePrefix = {arXiv},
       eprint = {2105.06042},
 primaryClass = {astro-ph.EP},
       adsurl = {https://ui.adsabs.harvard.edu/abs/2021ApJ...919..107L}
}

@ARTICLE{Ligterink2024NatAs...8.1257L,
       author = {{Ligterink}, Niels F.~W. and {Pinilla}, Paola and {van der Marel}, Nienke and {van Scheltinga}, Jeroen Terwisscha and {Booth}, Alice S. and {Alexander}, Conel M. O'D. and {Riebe}, My E.~I.},
        title = "{The rapid formation of macromolecules in irradiated ice of protoplanetary disk dust traps}",
      journal = {Nature Astronomy},
         year = 2024,
        month = oct,
       volume = {8},
       number = {10},
        pages = {1257-1263},
          doi = {10.1038/s41550-024-02334-4},
       adsurl = {https://ui.adsabs.harvard.edu/abs/2024NatAs...8.1257L}
}

@ARTICLE{Lim2024ApJ...969..130L,
       author = {{Lim}, Jeonghoon and {Simon}, Jacob B. and {Li}, Rixin and {Armitage}, Philip J. and {Carrera}, Daniel and {Lyra}, Wladimir and {Rea}, David G. and {Yang}, Chao-Chin and {Youdin}, Andrew N.},
        title = "{Streaming Instability and Turbulence: Conditions for Planetesimal Formation}",
      journal = {\apj},
         year = 2024,
        month = jul,
       volume = {969},
       number = {2},
          eid = {130},
        pages = {130},
          doi = {10.3847/1538-4357/ad47a2},
archivePrefix = {arXiv},
       eprint = {2312.12508},
 primaryClass = {astro-ph.EP},
       adsurl = {https://ui.adsabs.harvard.edu/abs/2024ApJ...969..130L}
}

@ARTICLE{Lim2025ApJ...981..160L,
       author = {{Lim}, Jeonghoon and {Simon}, Jacob B. and {Li}, Rixin and {Carrera}, Daniel and {Baronett}, Stanley A. and {Youdin}, Andrew N. and {Lyra}, Wladimir and {Yang}, Chao-Chin},
        title = "{Probing Conditions for Strong Clumping by the Streaming Instability: Small Dust Grains and Low Dust-to-gas Density Ratio}",
      journal = {\apj},
         year = 2025,
        month = mar,
       volume = {981},
       number = {2},
          eid = {160},
        pages = {160},
          doi = {10.3847/1538-4357/adb311},
archivePrefix = {arXiv},
       eprint = {2410.17319},
 primaryClass = {astro-ph.EP},
       adsurl = {https://ui.adsabs.harvard.edu/abs/2025ApJ...981..160L}
}

@ARTICLE{Lim2025ApJ...993...12L,
       author = {{Lim}, Jeonghoon and {Baronett}, Stanley A. and {Simon}, Jacob B. and {Yang}, Chao-Chin and {Sengupta}, Debanjan and {Umurhan}, Orkan M. and {Lyra}, Wladimir},
        title = "{Bridging Unstratified and Stratified Simulations of the Streaming Instability for {\ensuremath{\tau}}$_{s}$ = 0.1 Grains}",
      journal = {\apj},
         year = 2025,
        month = nov,
       volume = {993},
       number = {1},
          eid = {12},
        pages = {12},
          doi = {10.3847/1538-4357/ae01a6},
archivePrefix = {arXiv},
       eprint = {2505.23902},
 primaryClass = {astro-ph.EP},
       adsurl = {https://ui.adsabs.harvard.edu/abs/2025ApJ...993...12L}
}

@ARTICLE{Lin2015ApJ...811...17L,
       author = {{Lin}, Min-Kai and {Youdin}, Andrew N.},
        title = "{Cooling Requirements for the Vertical Shear Instability in Protoplanetary Disks}",
      journal = {\apj},
         year = 2015,
        month = sep,
       volume = {811},
       number = {1},
          eid = {17},
        pages = {17},
          doi = {10.1088/0004-637X/811/1/17},
archivePrefix = {arXiv},
       eprint = {1505.02163},
 primaryClass = {astro-ph.EP},
       adsurl = {https://ui.adsabs.harvard.edu/abs/2015ApJ...811...17L}
}

@ARTICLE{Lin2017ApJ...849..129L,
       author = {{Lin}, Min-Kai and {Youdin}, Andrew N.},
        title = "{A Thermodynamic View of Dusty Protoplanetary Disks}",
      journal = {\apj},
         year = 2017,
        month = nov,
       volume = {849},
       number = {2},
          eid = {129},
        pages = {129},
          doi = {10.3847/1538-4357/aa92cd},
archivePrefix = {arXiv},
       eprint = {1708.02945},
 primaryClass = {astro-ph.EP},
       adsurl = {https://ui.adsabs.harvard.edu/abs/2017ApJ...849..129L}
}

@ARTICLE{Lin2019MNRAS.485.5221L,
       author = {{Lin}, Min-Kai},
        title = "{Dust settling against hydrodynamic turbulence in protoplanetary discs}",
      journal = {\mnras},
         year = 2019,
        month = jun,
       volume = {485},
       number = {4},
        pages = {5221-5234},
          doi = {10.1093/mnras/stz701},
archivePrefix = {arXiv},
       eprint = {1903.03620},
 primaryClass = {astro-ph.EP},
       adsurl = {https://ui.adsabs.harvard.edu/abs/2019MNRAS.485.5221L}
}

@ARTICLE{Lin2021ApJ...907...64L,
       author = {{Lin}, Min-Kai},
        title = "{Stratified and Vertically Shearing Streaming Instabilities in Protoplanetary Disks}",
      journal = {\apj},
         year = 2021,
        month = feb,
       volume = {907},
       number = {2},
          eid = {64},
        pages = {64},
          doi = {10.3847/1538-4357/abcd9b},
archivePrefix = {arXiv},
       eprint = {2011.12300},
 primaryClass = {astro-ph.EP},
       adsurl = {https://ui.adsabs.harvard.edu/abs/2021ApJ...907...64L}
}

@ARTICLE{Matthijsse2025A&A...695A.158M,
       author = {{Matthijsse}, Jip and {Aly}, Hossam and {Paardekooper}, Sijme-Jan},
        title = "{Polydisperse formation of planetesimals: The dust size distribution in clumps}",
      journal = {\aap},
         year = 2025,
        month = mar,
       volume = {695},
          eid = {A158},
        pages = {A158},
          doi = {10.1051/0004-6361/202453072},
archivePrefix = {arXiv},
       eprint = {2502.01752},
 primaryClass = {astro-ph.EP},
       adsurl = {https://ui.adsabs.harvard.edu/abs/2025A&A...695A.158M}
}

@ARTICLE{MatthijsseInprepA&APaperII,
       author = {{Matthijsse}, Jip and {Aly}, Hossam and {Paardekooper}, Sijme-Jan},
        title = "{Dust-driven vertical shear instability in a local and isothermal shearing box - II. Vertical mixing of dust and the stabilizing effect of dust feedback}",
         year ={In prep.},
    shorthand = {Paper II}
}

@ARTICLE{McNally2015ApJ...811..121M,
       author = {{McNally}, Colin P. and {Pessah}, Martin E.},
        title = "{On Vertically Global, Horizontally Local Models for Astrophysical Disks}",
      journal = {\apj},
         year = 2015,
        month = oct,
       volume = {811},
       number = {2},
          eid = {121},
        pages = {121},
          doi = {10.1088/0004-637X/811/2/121},
archivePrefix = {arXiv},
       eprint = {1406.4864},
 primaryClass = {astro-ph.EP},
       adsurl = {https://ui.adsabs.harvard.edu/abs/2015ApJ...811..121M}
}

@ARTICLE{McNally2021MNRAS.502.1469M,
       author = {{McNally}, Colin P. and {Lovascio}, Francesco and {Paardekooper}, Sijme-Jan},
        title = "{Polydisperse streaming instability - III. Dust evolution encourages fast instability}",
      journal = {\mnras},
         year = 2021,
        month = mar,
       volume = {502},
       number = {1},
        pages = {1469-1486},
          doi = {10.1093/mnras/stab112},
archivePrefix = {arXiv},
       eprint = {2101.04761},
 primaryClass = {astro-ph.EP},
       adsurl = {https://ui.adsabs.harvard.edu/abs/2021MNRAS.502.1469M}
}

@ARTICLE{MelonFuksman2024A&A...682A.140M,
       author = {{Melon Fuksman}, Julio David and {Flock}, Mario and {Klahr}, Hubert},
        title = "{Vertical shear instability in two-moment radiation-hydrodynamical simulations of irradiated protoplanetary disks. II. Secondary instabilities and stability regions}",
      journal = {\aap},
         year = 2024,
        month = feb,
       volume = {682},
          eid = {A140},
        pages = {A140},
          doi = {10.1051/0004-6361/202346555},
archivePrefix = {arXiv},
       eprint = {2312.06890},
 primaryClass = {astro-ph.EP},
       adsurl = {https://ui.adsabs.harvard.edu/abs/2024A&A...682A.140M}
}

@ARTICLE{Nakagawa1986Icar...67..375N,
       author = {{Nakagawa}, Y. and {Sekiya}, M. and {Hayashi}, C.},
        title = "{Settling and growth of dust particles in a laminar phase of a low-mass solar nebula}",
      journal = {\icarus},
         year = 1986,
        month = sep,
       volume = {67},
       number = {3},
        pages = {375-390},
          doi = {10.1016/0019-1035(86)90121-1},
       adsurl = {https://ui.adsabs.harvard.edu/abs/1986Icar...67..375N}
}

@ARTICLE{Nelson2013MNRAS.435.2610N,
       author = {{Nelson}, Richard P. and {Gressel}, Oliver and {Umurhan}, Orkan M.},
        title = "{Linear and non-linear evolution of the vertical shear instability in accretion discs}",
      journal = {\mnras},
         year = 2013,
        month = nov,
       volume = {435},
       number = {3},
        pages = {2610-2632},
          doi = {10.1093/mnras/stt1475},
archivePrefix = {arXiv},
       eprint = {1209.2753},
 primaryClass = {astro-ph.EP},
       adsurl = {https://ui.adsabs.harvard.edu/abs/2013MNRAS.435.2610N}
}

@ARTICLE{Paardekooper2020MNRAS.499.4223P,
       author = {{Paardekooper}, Sijme-Jan and {McNally}, Colin P. and {Lovascio}, Francesco},
        title = "{Polydisperse streaming instability - I. Tightly coupled particles and the terminal velocity approximation}",
      journal = {\mnras},
         year = 2020,
        month = dec,
       volume = {499},
       number = {3},
        pages = {4223-4238},
          doi = {10.1093/mnras/staa3162},
archivePrefix = {arXiv},
       eprint = {2010.01145},
 primaryClass = {astro-ph.EP},
       adsurl = {https://ui.adsabs.harvard.edu/abs/2020MNRAS.499.4223P}
}

@ARTICLE{Paardekooper2021MNRAS.502.1579P,
       author = {{Paardekooper}, Sijme-Jan and {McNally}, Colin P. and {Lovascio}, Francesco},
        title = "{Polydisperse streaming instability - II. Methods for solving the linear stability problem}",
      journal = {\mnras},
         year = 2021,
        month = apr,
       volume = {502},
       number = {2},
        pages = {1579-1595},
          doi = {10.1093/mnras/stab111},
archivePrefix = {arXiv},
       eprint = {2101.04763},
 primaryClass = {astro-ph.EP},
       adsurl = {https://ui.adsabs.harvard.edu/abs/2021MNRAS.502.1579P}
}

@ARTICLE{Pfeil2023ApJ...959..121P,
       author = {{Pfeil}, Thomas and {Birnstiel}, Tilman and {Klahr}, Hubert},
        title = "{Dust Coagulation Reconciles Protoplanetary Disk Observations with the Vertical Shear Instability. I. Dust Coagulation and the VSI Dead Zone}",
      journal = {\apj},
         year = 2023,
        month = dec,
       volume = {959},
       number = {2},
          eid = {121},
        pages = {121},
          doi = {10.3847/1538-4357/ad00af},
archivePrefix = {arXiv},
       eprint = {2310.07332},
 primaryClass = {astro-ph.EP},
       adsurl = {https://ui.adsabs.harvard.edu/abs/2023ApJ...959..121P}
}

@ARTICLE{Pfeil2024A&A...687L...5P,
       author = {{Pfeil}, Thomas and {Birnstiel}, Til and {Klahr}, Hubert},
        title = "{Vertical shear instability with dust evolution and consistent cooling times. On the importance of the initial dust distribution}",
      journal = {\aap},
         year = 2024,
        month = jul,
       volume = {687},
          eid = {L5},
        pages = {L5},
          doi = {10.1051/0004-6361/202449323},
archivePrefix = {arXiv},
       eprint = {2406.10335},
 primaryClass = {astro-ph.EP},
       adsurl = {https://ui.adsabs.harvard.edu/abs/2024A&A...687L...5P}
}

@ARTICLE{Schafer2020A&A...635A.190S,
       author = {{Sch{\"a}fer}, Urs and {Johansen}, Anders and {Banerjee}, Robi},
        title = "{The coexistence of the streaming instability and the vertical shear instability in protoplanetary disks}",
      journal = {\aap},
         year = 2020,
        month = mar,
       volume = {635},
          eid = {A190},
        pages = {A190},
          doi = {10.1051/0004-6361/201937371},
archivePrefix = {arXiv},
       eprint = {2002.07185},
 primaryClass = {astro-ph.EP},
       adsurl = {https://ui.adsabs.harvard.edu/abs/2020A&A...635A.190S}
}

@ARTICLE{Schafer2025A&A...694A..57S,
       author = {{Sch{\"a}fer}, Urs and {Johansen}, Anders and {Flock}, Mario},
        title = "{The coexistence of the streaming instability and the vertical shear instability in protoplanetary disks: Scale dependence of dust diffusion}",
      journal = {\aap},
         year = 2025,
        month = feb,
       volume = {694},
          eid = {A57},
        pages = {A57},
          doi = {10.1051/0004-6361/202453439},
archivePrefix = {arXiv},
       eprint = {2501.07633},
 primaryClass = {astro-ph.EP},
       adsurl = {https://ui.adsabs.harvard.edu/abs/2025A&A...694A..57S}
}

@ARTICLE{Semenov2006ApJ...647L..57S,
       author = {{Semenov}, D. and {Wiebe}, D. and {Henning}, Th.},
        title = "{Gas-Phase CO in Protoplanetary Disks: A Challenge for Turbulent Mixing}",
      journal = {\apjl},
         year = 2006,
        month = aug,
       volume = {647},
       number = {1},
        pages = {L57-L60},
          doi = {10.1086/507096},
archivePrefix = {arXiv},
       eprint = {astro-ph/0607010},
 primaryClass = {astro-ph},
       adsurl = {https://ui.adsabs.harvard.edu/abs/2006ApJ...647L..57S}
}

@ARTICLE{semenov2011ApJS..196...25S,
       author = {{Semenov}, D. and {Wiebe}, D.},
        title = "{Chemical Evolution of Turbulent Protoplanetary Disks and the Solar Nebula}",
      journal = {\apjs},
         year = 2011,
        month = oct,
       volume = {196},
       number = {2},
          eid = {25},
        pages = {25},
          doi = {10.1088/0067-0049/196/2/25},
archivePrefix = {arXiv},
       eprint = {1104.4358},
 primaryClass = {astro-ph.GA},
       adsurl = {https://ui.adsabs.harvard.edu/abs/2011ApJS..196...25S}
}

@ARTICLE{Sengupta2023ApJ...942...74S,
       author = {{Sengupta}, Debanjan and {Umurhan}, Orkan M.},
        title = "{Turbulence in Particle-laden Midplane Layers of Planet-forming Disks}",
      journal = {\apj},
         year = 2023,
        month = jan,
       volume = {942},
       number = {2},
          eid = {74},
        pages = {74},
          doi = {10.3847/1538-4357/ac9411},
archivePrefix = {arXiv},
       eprint = {2209.11205},
 primaryClass = {astro-ph.EP},
       adsurl = {https://ui.adsabs.harvard.edu/abs/2023ApJ...942...74S}
}

@ARTICLE{Shakura1973A&A....24..337S,
       author = {{Shakura}, N.~I. and {Sunyaev}, R.~A.},
        title = "{Black holes in binary systems. Observational appearance.}",
      journal = {\aap},
         year = 1973,
        month = jan,
       volume = {24},
        pages = {337-355},
       adsurl = {https://ui.adsabs.harvard.edu/abs/1973A&A....24..337S}
}

@ARTICLE{Shtemler2019MNRAS.488.4207S,
       author = {{Shtemler}, Yuri and {Mond}, Michael},
        title = "{Vertical shear-induced resonant triads in Keplerian discs}",
      journal = {\mnras},
         year = 2019,
        month = sep,
       volume = {488},
       number = {3},
        pages = {4207-4219},
          doi = {10.1093/mnras/stz2032},
       adsurl = {https://ui.adsabs.harvard.edu/abs/2019MNRAS.488.4207S}
}

@ARTICLE{Shtemler2020MNRAS.499.3222S,
       author = {{Shtemler}, Yuri and {Mond}, Michael},
        title = "{Clustering of the resonant triads induced by vertical-shear instability in astrophysical discs}",
      journal = {\mnras},
         year = 2020,
        month = dec,
       volume = {499},
       number = {3},
        pages = {3222-3232},
          doi = {10.1093/mnras/staa3069},
       adsurl = {https://ui.adsabs.harvard.edu/abs/2020MNRAS.499.3222S}
}

@ARTICLE{Shtemler2025MNRAS.543.1313S,
       author = {{Shtemler}, Yuri and {Mond}, Michael},
        title = "{The vertical shear instability in turbulent protoplanetary discs. Marginal stability and kinematic analysis of resonant triads}",
      journal = {\mnras},
         year = 2025,
        month = oct,
       volume = {543},
       number = {2},
        pages = {1313-1321},
          doi = {10.1093/mnras/staf1573},
       adsurl = {https://ui.adsabs.harvard.edu/abs/2025MNRAS.543.1313S}
}

@ARTICLE{Stoll2014A&A...572A..77S,
       author = {{Stoll}, Moritz H.~R. and {Kley}, Wilhelm},
        title = "{Vertical shear instability in accretion disc models with radiation transport}",
      journal = {\aap},
         year = 2014,
        month = dec,
       volume = {572},
          eid = {A77},
        pages = {A77},
          doi = {10.1051/0004-6361/201424114},
archivePrefix = {arXiv},
       eprint = {1409.8429},
 primaryClass = {astro-ph.EP},
       adsurl = {https://ui.adsabs.harvard.edu/abs/2014A&A...572A..77S}
}

@ARTICLE{Stoll2017A&A...599L...6S,
       author = {{Stoll}, Moritz H.~R. and {Kley}, Wilhelm and {Picogna}, Giovanni},
        title = "{Anisotropic hydrodynamic turbulence in accretion disks}",
      journal = {\aap},
         year = 2017,
        month = mar,
       volume = {599},
          eid = {L6},
        pages = {L6},
          doi = {10.1051/0004-6361/201630226},
archivePrefix = {arXiv},
       eprint = {1702.00334},
 primaryClass = {astro-ph.SR},
       adsurl = {https://ui.adsabs.harvard.edu/abs/2017A&A...599L...6S}
}

@ARTICLE{Squire2018ApJ...856L..15S,
       author = {{Squire}, J. and {Hopkins}, P.~F.},
        title = "{Resonant Drag Instability of Grains Streaming in Fluids}",
      journal = {\apjl},
         year = 2018,
        month = mar,
       volume = {856},
       number = {1},
          eid = {L15},
        pages = {L15},
          doi = {10.3847/2041-8213/aab54d},
archivePrefix = {arXiv},
       eprint = {1706.05020},
 primaryClass = {astro-ph.EP},
       adsurl = {https://ui.adsabs.harvard.edu/abs/2018ApJ...856L..15S}
}

@ARTICLE{Squire2018MNRAS.477.5011S,
       author = {{Squire}, Jonathan and {Hopkins}, Philip F.},
        title = "{Resonant drag instabilities in protoplanetary discs: the streaming instability and new, faster growing instabilities}",
      journal = {\mnras},
         year = 2018,
        month = jul,
       volume = {477},
       number = {4},
        pages = {5011-5040},
          doi = {10.1093/mnras/sty854},
archivePrefix = {arXiv},
       eprint = {1711.03975},
 primaryClass = {astro-ph.EP},
       adsurl = {https://ui.adsabs.harvard.edu/abs/2018MNRAS.477.5011S}
}

@ARTICLE{Tatarelli2026A&A...711A.252T,
       author = {{Tatarelli}, M. and {Morbidelli}, A. and {Lega}, E.},
        title = "{Dust dynamics in disk dust traps and late planetesimal formation}",
      journal = {\aap},
         year = 2026,
        month = jul,
       volume = {711},
          eid = {A252},
        pages = {A252},
          doi = {10.1051/0004-6361/202659196},
archivePrefix = {arXiv},
       eprint = {2605.29981},
 primaryClass = {astro-ph.EP},
       adsurl = {https://ui.adsabs.harvard.edu/abs/2026A&A...711A.252T}
}

@ARTICLE{Temmink2025A&A...693A.101T,
       author = {{Temmink}, Milou and {Booth}, Alice S. and {Leemker}, Margot and {van der Marel}, Nienke and {van Dishoeck}, Ewine F. and {Evans}, Lucy and {Keyte}, Luke and {Law}, Charles J. and {Notsu}, Shota and {{\"O}berg}, Karin and {Walsh}, Catherine},
        title = "{Characterising the molecular line emission in the asymmetric Oph-IRS 48 dust trap: Temperatures, timescales, and sub-thermal excitation}",
      journal = {\aap},
         year = 2025,
        month = jan,
       volume = {693},
          eid = {A101},
        pages = {A101},
          doi = {10.1051/0004-6361/202452175},
archivePrefix = {arXiv},
       eprint = {2411.12418},
 primaryClass = {astro-ph.EP},
       adsurl = {https://ui.adsabs.harvard.edu/abs/2025A&A...693A.101T}
}

@ARTICLE{Urpin1998MNRAS.294..399U,
       author = {{Urpin}, V. and {Brandenburg}, A.},
        title = "{Magnetic and vertical shear instabilities in accretion discs}",
      journal = {\mnras},
         year = 1998,
        month = mar,
       volume = {294},
       number = {3},
        pages = {399-406},
          doi = {10.1046/j.1365-8711.1998.01118.x10.1111/j.1365-8711.1998.01118.x},
       adsurl = {https://ui.adsabs.harvard.edu/abs/1998MNRAS.294..399U}
}

@ARTICLE{Urpin2003A&A...404..397U,
       author = {{Urpin}, V.},
        title = "{A comparison study of the vertical and magnetic shear instabilities in accretion discs}",
      journal = {\aap},
         year = 2003,
        month = jun,
       volume = {404},
        pages = {397-403},
          doi = {10.1051/0004-6361:20030513},
       adsurl = {https://ui.adsabs.harvard.edu/abs/2003A&A...404..397U}
}

@ARTICLE{vandermarel_A_2021A&A...651L...5V,
       author = {{van der Marel}, Nienke and {Booth}, Alice S. and {Leemker}, Margot and {van Dishoeck}, Ewine F. and {Ohashi}, Satoshi},
        title = "{A major asymmetric ice trap in a planet-forming disk. I. Formaldehyde and methanol}",
      journal = {\aap},
         year = 2021,
        month = jul,
       volume = {651},
          eid = {L5},
        pages = {L5},
          doi = {10.1051/0004-6361/202141051},
archivePrefix = {arXiv},
       eprint = {2104.08906},
 primaryClass = {astro-ph.EP},
       adsurl = {https://ui.adsabs.harvard.edu/abs/2021A&A...651L...5V}
}

@ARTICLE{Virtanen2020NatMe..17..261V,
       author = {{Virtanen}, Pauli and {Gommers}, Ralf and {Oliphant}, Travis E. and {Haberland}, Matt and {Reddy}, Tyler and {Cournapeau}, David and {Burovski}, Evgeni and {Peterson}, Pearu and {Weckesser}, Warren and {Bright}, Jonathan and {van der Walt}, St{\'e}fan J. and {Brett}, Matthew and {Wilson}, Joshua and {Millman}, K. Jarrod and {Mayorov}, Nikolay and {Nelson}, Andrew R.~J. and {Jones}, Eric and {Kern}, Robert and {Larson}, Eric and {Carey}, C.~J. and {Polat}, {\.I}lhan and {Feng}, Yu and {Moore}, Eric W. and {VanderPlas}, Jake and {Laxalde}, Denis and {Perktold}, Josef and {Cimrman}, Robert and {Henriksen}, Ian and {Quintero}, E.~A. and {Harris}, Charles R. and {Archibald}, Anne M. and {Ribeiro}, Ant{\^o}nio H. and {Pedregosa}, Fabian and {van Mulbregt}, Paul and {SciPy 1. 0 Contributors}},
        title = "{SciPy 1.0: fundamental algorithms for scientific computing in Python}",
      journal = {Nature Methods},
         year = 2020,
        month = feb,
       volume = {17},
        pages = {261-272},
          doi = {10.1038/s41592-019-0686-2},
archivePrefix = {arXiv},
       eprint = {1907.10121},
 primaryClass = {cs.MS},
       adsurl = {https://ui.adsabs.harvard.edu/abs/2020NatMe..17..261V}
}

@ARTICLE{Woitke2022A&A...668A.164W,
       author = {{Woitke}, P. and {Arabhavi}, A.~M. and {Kamp}, I. and {Thi}, W.-F.},
        title = "{Mixing and diffusion in protoplanetary disc chemistry}",
      journal = {\aap},
         year = 2022,
        month = dec,
       volume = {668},
          eid = {A164},
        pages = {A164},
          doi = {10.1051/0004-6361/202244554},
archivePrefix = {arXiv},
       eprint = {2209.12233},
 primaryClass = {astro-ph.EP},
       adsurl = {https://ui.adsabs.harvard.edu/abs/2022A&A...668A.164W}
}

@ARTICLE{Youdin2005ApJ...620..459Y,
       author = {{Youdin}, Andrew N. and {Goodman}, Jeremy},
        title = "{Streaming Instabilities in Protoplanetary Disks}",
      journal = {\apj},
         year = 2005,
        month = feb,
       volume = {620},
       number = {1},
        pages = {459-469},
          doi = {10.1086/426895},
archivePrefix = {arXiv},
       eprint = {astro-ph/0409263},
 primaryClass = {astro-ph},
       adsurl = {https://ui.adsabs.harvard.edu/abs/2005ApJ...620..459Y}
}

@ARTICLE{Youdin2007Icar..192..588Y,
       author = {{Youdin}, Andrew N. and {Lithwick}, Yoram},
        title = "{Particle stirring in turbulent gas disks: Including orbital oscillations}",
      journal = {\icarus},
         year = 2007,
        month = dec,
       volume = {192},
       number = {2},
        pages = {588-604},
          doi = {10.1016/j.icarus.2007.07.012},
archivePrefix = {arXiv},
       eprint = {0707.2975},
 primaryClass = {astro-ph},
       adsurl = {https://ui.adsabs.harvard.edu/abs/2007Icar..192..588Y}
}

@misc{DHPC2024,
author = {{D}elft {H}igh {P}erformance {C}omputing {C}entre ({DHPC})},
title = {{D}elft{B}lue {S}upercomputer ({P}hase 2)},
year = {2024},
howpublished = {\url{https://www.tudelft.nl/dhpc/ark:/44463/DelftBluePhase2}},
ark = {ark:/44463/DelftBluePhase2}
}


\begin{appendix}
\onecolumn
\section{Convergence with resolution heatmaps}

\begin{figure*}[ht!]
   \resizebox{\hsize}{!}
            {\includegraphics {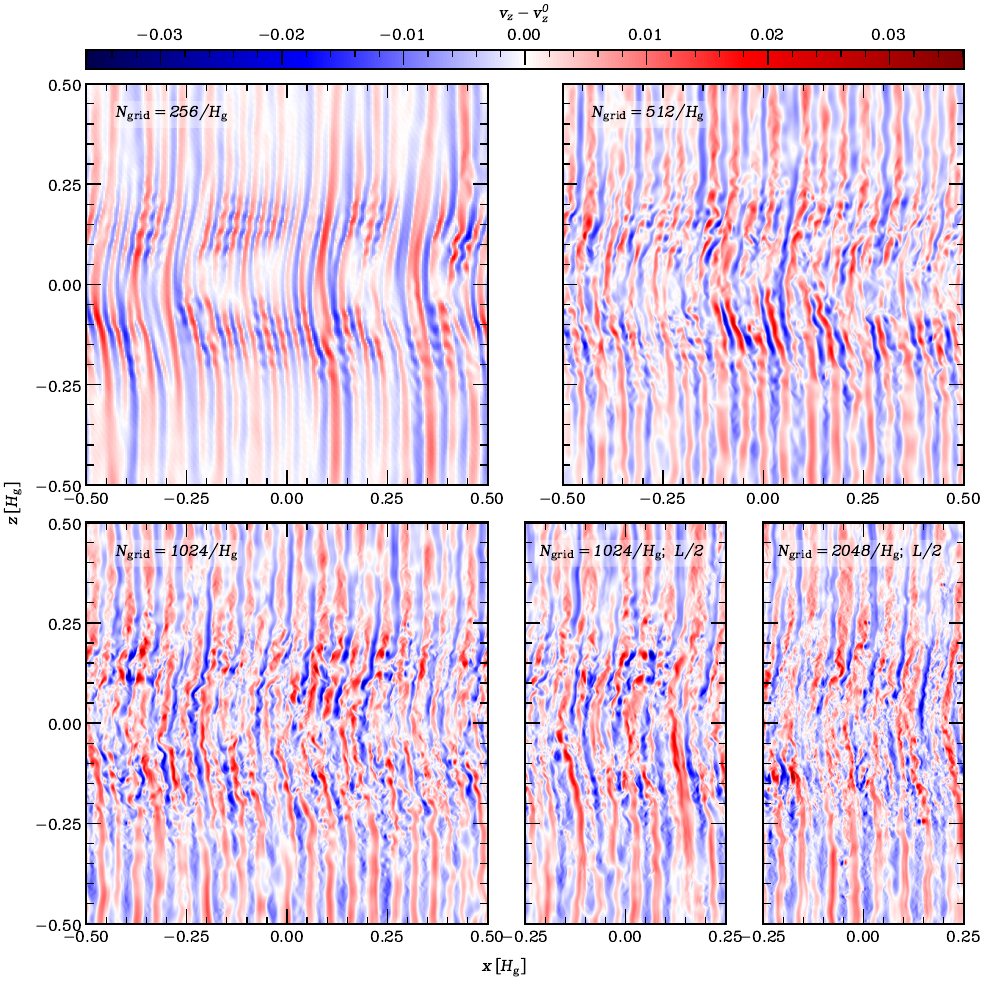}}
    \centering
    \caption{Heatmap of 5 runs with different resolution and box sizes, run \texttt{DVSI}$_{N,\, 256}$ (top left), run \texttt{DVSI}$_{N,\, 512}$ (top right) and run \texttt{DVSI} (bottom left), and the two runs with half the box size \texttt{DVSI*}$_{N,\, 1024}$ and \texttt{DVSI*}$_{N,\, 2048}$ (bottom right), showing a snapshot of the perturbed vertical velocity of the gas at $\Omega t=500$. The heatmaps do not cover the full vertical extent of the simulation of $z_\mathrm{max}=4H_\mathrm{g}$, showing $L_z/8$ for the first three heatmaps and the last two runs $z_\mathrm{max}=2H_\mathrm{g}$, only showing $L_z/4$.}
    \label{fig:DVSI_Ngrid_t500}
\end{figure*}

\twocolumn

\onecolumn
\section{Numerical artefacts in FARGO3D}

\begin{figure*}[ht!]
   \resizebox{0.75\hsize}{!}
            {\includegraphics {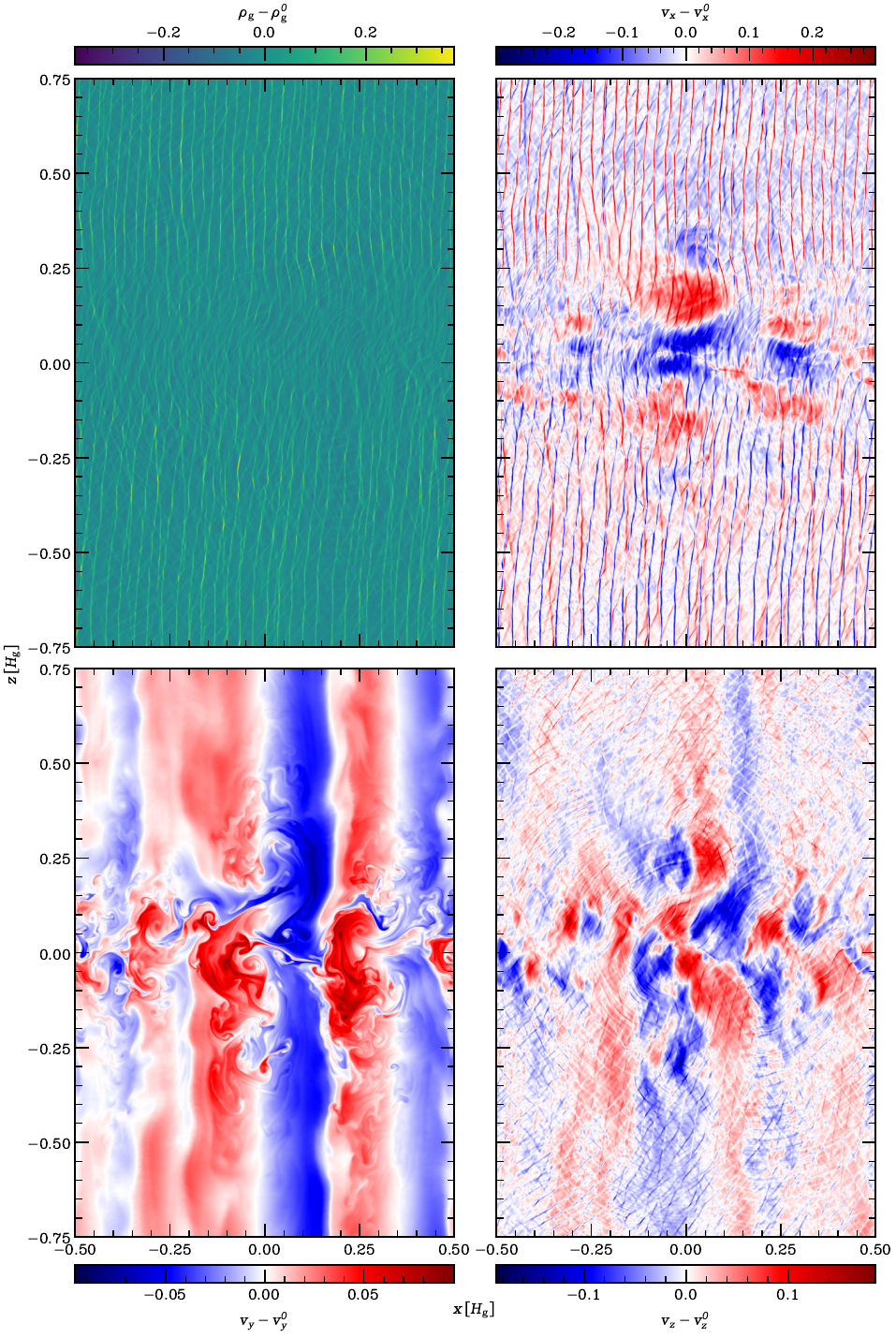}}
    \centering        
    \caption{Heatmap of run \texttt{DVSI}$_{\alpha_D,\, 1\mathrm{e}-6}$ at $\Omega t = 250$, showing the perturbed densities and velocities in all directions, where the background densities and velocities are subtracted. }
    \label{fig:DVSI_gas_t500_ad6}
\end{figure*}
\twocolumn

\end{appendix}

\end{document}